\documentclass[preprint]{vgtc}               

\graphicspath{{figures/}{pictures/}{images/}{./}} 

\usepackage{times}                     

\usepackage{tabu}                      
\usepackage{booktabs}                  
\usepackage{lipsum}                    
\usepackage{mwe}                       

\usepackage{enumerate}
\usepackage{cool}

\newcommand{\mySetNotation}[1]{{\mathbb{#1}}}
\newcommand{\RRSet}{{\mySetNotation{R}}}

\newcommand{\myBoldNotation}[1]{{\mathbf #1}}

\newcommand{\vp}   {{\myBoldNotation{p}}}

\newcommand{\vx}   {{\myBoldNotation{x}}}

\newcommand{\mA}   {{\myBoldNotation{A}}}

\newcommand{\mI}   {{\myBoldNotation{I}}}

\newcommand{\mT}   {{\myBoldNotation{T}}}

\Style{IdentityMatrixSymb=\mI,%
       DSymb={\mathrm d},%
       IntegrateDifferentialDSymb={\mathrm d},%
       TrParen=p,%
       DetParen=p}
\Style{DDisplayFunc=outset}

\usepackage{mathptmx}                  

\DeclareUnicodeCharacter{3B1}{\ensuremath{\alpha}}
\DeclareMathOperator{\mode}{mode}

\onlineid{0}

\vgtccategory{Research}

\vgtcinsertpkg

\usepackage{tikz}
\newcommand{\tikzcircle}[2][red,fill=red]{\tikz[baseline=-0.5ex]\draw[#1,radius=#2] (0,0) circle ;}%

\definecolor{mylight}{HTML}{AEBB9E}
\definecolor{mymid}{HTML}{FFE75C}
\definecolor{mydark}{HTML}{F0B100}

\ieeedoi{10.1109/UncertaintyVisualization63963.2024.00005}

\title{Exploring Uncertainty Visualization for Degenerate Tensors in 3D Symmetric
Second-Order Tensor Field Ensembles}

\author{Tadea Schmitz\thanks{e-mail: schmitz@cs.uni-koeln.de}\\ %
        \scriptsize RWTH Aachen University, Germany\\
        \scriptsize University of Cologne, Germany
\and Tim Gerrits\thanks{e-mail: gerrits@vis.rwth-aachen.de}\\ %
     \scriptsize RWTH Aachen University, Germany}

\teaser{
  \centering
  \includegraphics[width=\linewidth]{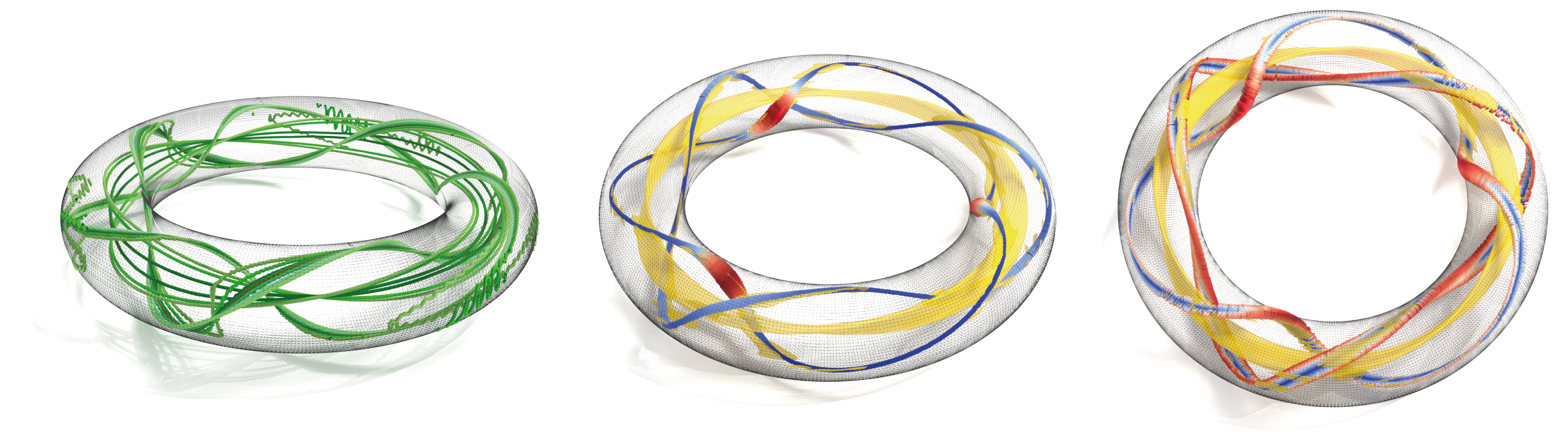}
  \caption{Ensemble of stress in an O-ring simulation with eight members and varying parameter $\alpha$, which amplifies anisotropy magnitude in the simulation. Left: Spaghetti plot visualization of all extracted degenerate tensor lines of ensemble members. Each member is represented by its own shape of green. Center: Uncertainty visualization using an enhanced meanLine showing mode standard deviation using radius and color as well as a probabilityBand in yellow. Right: Same probabilityBand visualization alongside a modeTube visualization indicating the spatial distribution of tensor mode values through color and shape.}
  \label{fig:teaser}
}

\abstract{
    Symmetric second-order tensors are fundamental in various scientific and engineering domains, as they can represent properties such as material stresses or diffusion processes in brain tissue.
    In recent years, several approaches have been introduced and improved to analyze these fields using topological features, such as degenerate tensor locations, i.e., the tensor has repeated eigenvalues, or normal surfaces.
    Traditionally, the identification of such features has been limited to single tensor fields.
    However, it has become common to create ensembles to account for uncertainties and variability in simulations and measurements.
    In this work, we explore novel methods for describing and visualizing degenerate tensor locations in 3D symmetric second-order tensor field ensembles.
    We base our considerations on the tensor mode and analyze its practicality in characterizing the uncertainty of degenerate tensor locations before proposing a variety of visualization strategies to effectively communicate degenerate tensor information.
    We demonstrate our techniques for synthetic and simulation data sets.
    The results indicate that the interplay of different descriptions for uncertainty can effectively convey information on degenerate tensor locations.
        
} 

\keywords{Second-Order Tensors, Symmetric Tensors, Tensor Topology, Degenerate Tensors, Uncertainty.}

\begin{document}


\firstsection{Introduction} \label{sec:introduction}

\maketitle

Tensors provide useful mathematical descriptions for complex physical phenomena in a large variety of domains including mechanical engineering, such as stress or strain, or medical applications, such as the diffusion in brain white matter, among others.
They thus appear frequently in scientific simulations that seek to recreate and describe these processes.
Summarizing tensor fields using topological features can provide an overview and quick insights into the underlying structure and behavior of those fields, which has led to numerous novel, fast, and robust techniques for the extraction and visualization of such features in recent years~\cite{palacios2015feature,roy2018robust,hung2023global}.
Of special interest, e.g., in mechanical engineering \cite{zhang2017applying, hergl2021visualization}, are degenerate tensor locations, i.e., locations where the spectral decomposition of a tensor yields two or three repeated eigenvalues, called double- or triple-degenerate tensors respectively.
However, most approaches assume tensor fields to be \textit{certain}, while real-world applications usually contain uncertainties in numerous steps of the processing pipeline~\cite{siddiqui2021uncertainty}.
Describing and visualizing uncertainty is listed as one of the big challenges in the analysis of tensor fields in the context of engineering applications~\cite{hlawitschka2014top}.
In simulation sciences, it is common to model uncertainty by means of ensemble simulations, i.e., repeated simulations with variations in their properties.
However, many visualization strategies used for single fields cannot be directly applied to ensembles, especially with a large number of members.
While there exist few approaches to communicate the uncertainty of tensor distributions using glyphs~\cite{jiao2012uncertainty, abbasloo2015visualizing, gerrits2019towards}, we are not aware of any techniques that provide similar tools for topological features such as degenerate tensor locations.

In this work, we develop visualization strategies that provide an effective overview of the overall behavior of tensor fields within an ensemble with a focus on degenerate tensor locations.
Assuming a normal distribution, one strategy is based on features extracted from the mean tensor field.
This leads to the definition of the \textit{meanLine}, which acts as a line-type representative of the whole ensemble and can further be enhanced with additional encoding of uncertainty, e.g., as \textit{modeTube}.
The second strategy is based on the distribution of tensor mode values within the ensemble, which we use to extract \textit{probabilityBands}, giving a global indicator of degenerate tensor locations.
We provide and discuss initial observations and evaluate the utility of these features using multiple ensembles of synthetic and simulated stress tensor fields.
The results indicate that our proposed strategies are simple yet effective in communicating uncertainty of degenerate tensor locations in ensembles on different levels of detail and represent a first step toward a more comprehensive analysis of uncertain tensor fields.

\section{Background} \label{sec:background}
To describe the uncertainty of degenerate tensors, we need to revisit relevant tensor properties.
This includes general tensor attributes, as well as descriptions of distributions representing uncertainty.
As we focus on three-dimensional symmetric second-order tensors, for the remainder of this work the term \textit{tensor} refers to this specific description.
A tensor $\mT$ represented by a symmetric $3 \times 3$ matrix can be decomposed into its orthogonal set of eigenvectors and corresponding real-valued eigenvalues $\lambda_i$ for $i \in [1,2,3]$.
By convention, $\lambda_1 \geq \lambda_2 \geq \lambda_3$, referred to as major, medium, and minor eigenvalues.
Tensor invariants provide means to describe the tensor independent of its frame of reference:
The sum of eigenvalues describes the trace $tr(\mT)$, while their product describes the determinant $|\mT|$.
Besides the spectral decomposition in eigenvectors and eigenvalues, the deviatoric decomposition allows to represent a tensor uniquely by a traceless tensor $\mA$ known as deviator, and a multiple of the identity matrix $\mI$, such that $\mA = \mT - \frac{tr(\mT)}{3}\mI$.
Another tensor invariant directly connected to tensor topology is the $\mode(\mT)$, which can be defined as
\begin{equation}\label{eq:mode}
\mode(\mT) = 3 \sqrt{6} \det \left(\frac{\mA}{|\mA|}\right)
\end{equation}
and provides a scalar value in the range $[-1, 1]$.\\

A tensor is considered \textit{degenerate} if it has at least two repeated eigenvalues~\cite{hesselink1997topology}, which leads to a directional discontinuity in the tensors' eigenvector field.
This feature can appear as double-degeneracy, i.e., $(\lambda_1 > \lambda_2 = \lambda_3)$ (linear) or $(\lambda_1 = \lambda_2 > \lambda_3)$ (planar), forming structurally stable lines known as \textit{degenerate tensor lines} in symmetric 3D tensor fields~\cite{zheng2005topological}, or triple-degeneracy, i.e., $(\lambda_1 = \lambda_2 = \lambda_3)$ (neutral) which is structurally unstable.
A linear degenerate tensor further exhibits a mode value of 1, a neutral degenerate tensor 0, and a planar degenerate tensor a mode value of -1.

A tensor field is a continuous function within a domain where for each location $\vx \in \RRSet^3$, a tensor $\mT(\vx)$ is assigned.
A linear tensor field is a special case~\cite{zhang2021degenerate}, where the tensor components are linear functions of their spatial coordinates and can be decomposed as 
\begin{equation}\label{eq:linField}
\mT(\vx) = \mT(x,y,z) = x \cdot \mathbf{T}_x + y \cdot \mathbf{T}_y + z \cdot \mathbf{T}_z +\mathbf{T}_0
\end{equation}
and $\mT_x, \mT_y, \mT_z$, and $\mT_0$ are 3D symmetric tensors.
We will use this definition later to define test cases with desired properties. 

An ensemble is a set of $m$ fields, such that for each location $\vx \in \RRSet^3$ within the domain, one obtains multiple tensors $\mT_i$ and $i \in [0,\dots,m]$.
The collection of tensors of the ensemble resembles the uncertainty of the tensor field.

Similar to previous work~\cite{zhang2017overview+, basser2007spectral, gerrits2019towards}, one can assume the ensemble members to follow Gaussian distribution, such that the likelihood of a random variable taking on a particular value is described in terms of their distribution function.
Thus, a component-wise mean tensor
\begin{equation}\label{meanTen}
    \Bar{\mT} = \frac{1}{m} \sum^{m}_{i=1}\mT_i
\end{equation}
 and a fourth-order covariance tensor $\Sigma$ can be used to describe the uncertainty for each location using
\begin{equation}
p(\mT) = \sqrt{\frac{1}{\sqrt{(2 \pi)^6 |\Sigma|}}} \exp \left( - \tfrac{1}{2} (\mT - \Bar{\mT}):\Sigma^{-1}:(\mT - \Bar{\mT}) \right).
\end{equation}
Similarly, a univariate normal distribution can be described by
\begin{equation}\label{eq:gauss}
p(x) = \frac{1}{\sqrt{2 \pi \sigma^2}} \exp \left( -\frac{(x - \mu)^2}{2 \sigma^2} \right)
\end{equation}
where $\mu$ is the mean value and $\sigma$ is the standard deviation.

\section{Related Works} \label{sec:relatedworks}
Our work seeks to introduce uncertainty visualization to the field of topological descriptions for symmetric second-order tensor fields.
We, therefore, first summarize contributions on general tensor visualization before looking into relevant works that cover these specific topics.

\subsection{Tensor Visualization}
Tensors are used in a variety of scientific domains, which has led to the development of independent tensor visualizations, often tailored toward a specific domain.
Especially the symmetric case, which frequently appears in mechanical engineering~\cite{hergl2021visualization} and the medical domain has seen various developments.
An extensive survey is provided by Bi et al.~\cite{bi2019survey}, who divide the visualization approaches for tensor fields into glyph-based techniques (see, e.g., ~\cite{schultz2010superquadric, gerrits2016glyphs}) and streamlines, which include fiber tracking~\cite{isenberg2015survey}, hyperstreamlines~\cite{fu2014topologically}, and, most relevant to us, topological features.
The survey of Kratz et al.~\cite{kratz2013visualization}, additionally classifies multi-view visualizations, which show different aspects of tensors combined~\cite{kratz2011visual, kretzschmar2023visual}.\\

The concept of topological structures and \textit{degenerate points} within 2D tensor fields is based on the work of Delmarcelle and Hesselink~\cite{delmarcelle1993visualizing}, which was later extended to the 3D case~\cite{hesselink1997topology}.
Zheng at al.~\cite{zheng2005topological, zheng2006degenerate} focused on the properties of degenerate tensors, resulting in the definition of degenerate curves and the development of several extraction strategies over the years improving in accuracy, robustness, and performance~\cite{tricoche2008invariant, palacios2015feature,roy2018robust}.
It was further shown that such curves are ridge and valley lines of the tensor mode~\cite{tricoche2008invariant, palacios2015feature}, which could additionally be used to extract features such as neutral surfaces~\cite{qu2020mode}.
Hung et al.~\cite{hung2023global} propose using graph-like structures to show the global topology of tensor fields including curves, surfaces, and their connections.
Oster et al.\cite{oster2018core} propose tensor core lines, a feature similar to vortex core lines, which indicates rotational behavior of eigenvectors, while Zobel and Scheuermann~\cite{zobel2018extremal} introduce the notion of extremal points using multiple invariants.
Finally, there are efforts to define similar features for non-symmetric tensor fields~\cite{hung2021feature}.

\subsection{Uncertainty Visualization}
Considering uncertainties and their visualization has become a top challenge in visualization research~\cite{potter2012quantification} and a general overview is provided by surveys~\cite{bonneau2014overview, padilla2020uncertainty, wang2018visualization}, domain-specific reports~\cite{weiskopf2022uncertainty}, or guides for the creation of analytic workflows~\cite{maack2024workflow}.
While a decade ago, only few works explored visualizations of uncertainties present in tensor fields~\cite{hlawitschka2014top, potter2012quantification}, there has been considerable work in this field.
Most notably, a variety of glyph visualizations have been proposed to indicate the impact of uncertainty on the tensor field.
These include comparative visualizations~\cite{zhang2015glyph, zhang2017comparative, meuschke2017glyph}, glyphs indicating variability of the major eigenvector directions~\cite{schultz2013hifive, jones2002spatial}, and glyphs that indicate shape and orientation variations of the representing tensor glyphs~\cite{gerrits2019towards, abbasloo2015visualizing, jiao2012uncertainty}.
Basser and Pajevic~\cite{basser2003normal, basser2007spectral} propose using radial glyphs to display information encoded in the covariance matrix, while Zhang et al.~\cite{zhang2017overview+} propose a framework combining several views of tensor properties independently.\\

For continuous methods, especially the analysis of diffusion tensor imaging (DTI) data has seen the incorporation of uncertainties~\cite{siddiqui2021uncertainty, jiao2010metrics}, e.g., by using bootstrapping techniques in fiber tracking~\cite{schultz2014fuzzy, siddiqui2021progressive}.
Pöthkow and Hege~\cite{pothkow2012uncertainty} propose the extraction of isosurfaces in fields describing fractional anisotropy and relative anisotropy, where eigenvalue uncertainty is propagated to the shape and position of the surfaces. 
Finally, there exist approaches that seek to capture the uncertainty of line-type features derived from scalar- or vector-valued fields. 
In the context of meteorological forecasting, Spaghetti plots~\cite{potter2009ensemble}, contour boxplots~\cite{whitaker2013contour}, and curve boxplots~\cite{mirzargar2014curve} are frequently used.
However, degenerate tensor lines suffer from similar restrictions as vortex core lines, namely a missing common parametrization.
While this can be addressed~\cite{chaves2024depth, zhang2022enconvis}, not all ensemble members might possess that specific feature~\cite{gerrits2018approximate}, which might be misguiding and lead to a misrepresentation of the ensemble behavior.\\

After revisiting the literature, it seems there exists a gap in techniques to visualize topological features in uncertain tensor fields.
In the following, we propose a first solution to this challenge.

\section{Mode of Tensor Ensemble}
\begin{figure}
    \centering
    \includegraphics[width=\linewidth]{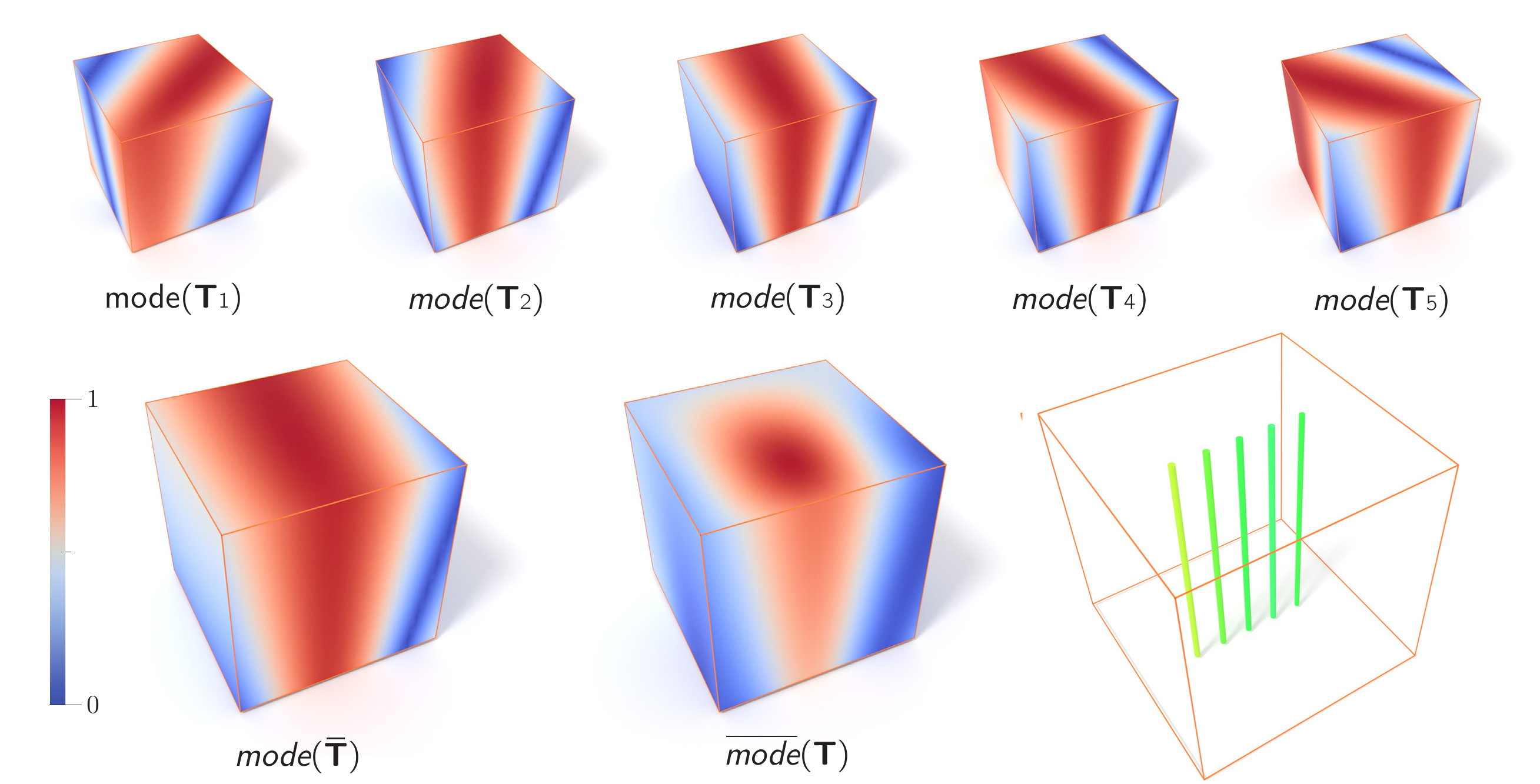}
    \caption{Tensor mode distribution in a synthetic ensemble. Top row: $\mode_{abs}$ values represented as color values using the cool-to-warm color map for the different members. Bottom left: $\mode_{abs}$ value of the mean tensor field. Bottom center: mean $\mode_{abs}$ of all members. Bottom right: spaghetti plot of extracted degenerate tensor lines.}
    \label{fig:modeCompare}
\end{figure}

On a conceptional level, our reasoning is simple:
The mode values of a single tensor field, as described by \cref{eq:mode}, can be used to extract degenerate tensor locations, thus, we seek to use the mode value of a tensor ensemble to do the same.
Yet, it is unclear, how the mode value of an ensemble is defined and what type of feature most accurately represents degenerate tensor locations.
There are two instinctive answers to the first question:
\begin{enumerate}
    \item The mode of an ensemble is the mode value of the mean tensor of all ensemble members.
    \item The mode of an ensemble is constructed from the mode values of all ensemble members.
\end{enumerate}
While one seems more straightforward than the other, in the following, we explore both descriptions and show their implications on possible features.
Finally, we show, that both views are needed to provide a comprehensive description of degenerate tensor locations and their uncertainties.

\subsubsection*{Mode of Mean Tensor as Ensemble Representative}
A common approach to represent an ensemble with a single representative for each location in the field is using the mean tensor~(see, e.g., \cite{gerrits2019towards, zhang2015glyph, zhang2017overview+}) as described by \cref{meanTen}.
The resulting tensor field can be treated as a single, certain field, providing a single tensor $\Bar{\mT}(\vx)$ per location, thus allowing the application of any standard tensor visualization method.
Accordingly, the representation of a mode value for a tensor ensemble can be given by $\mode(\Bar{\mT}(\vx))$.
Calculating the component-wise mean tensor uses linear interpolation of tensor components.
This, however, does not result in a similar interpolation of tensor invariants, such as the tensor mode~\cite{gahm2012linear, hotz2010tensor}.
Thus $\mode(\Bar{\mT}(\vx))$ does not necessarily represent the mean mode values of the ensemble members, which we address with our second description.

\subsubsection*{Mean Mode as Ensemble Representative}
As the tensor mode provides a direct description of degenerate tensor locations, we propose calculating a mean mode value for each location in the ensemble.
However, we make the following simplification:
While mode values representing planar and linear degenerate tensors are both located at the boundaries of the mode value range, we are only interested in the occurrence of a tensor dual-degeneracy, regardless of its classification.
Due to their instability, we exclude neutral degenerate tensors from our considerations.
We propose using $\mode_{abs}(\mT) = |\mode(\mT)|$ with $\mode_{abs} \in [0,1]$.
Therefore, a mode value approaching $1$ indicates a tensor getting closer to becoming degenerate.
A value representing the tensor mode of the whole ensemble at a given location $\vx$ can therefore be provided by the mean of the mode values of all $m$ members:
\begin{equation}\label{meanMode}
    \overline{\mode}(\vx) = \frac{1}{m}\sum_{i = 1}^m \mode_{abs}(\mT_i(\vx)).
\end{equation}
This further allows us to describe the distribution of mode values using statistical tools as described in \cref{sec:background}.

\section{Ensemble Feature Definitions}
Using these considerations, we can now make use of two representations describing the mode values within a tensor ensemble.
A simple visual comparison of both using a color map shows significant differences between both fields as displayed in the bottom row of \cref{fig:modeCompare}.
However, both can be used to define new features for tensor field ensembles, which we designed with the following goals in mind:
\begin{enumerate}[i]
    \item The feature should be able to effectively summarize the behavior of degenerate tensor locations of the whole ensemble.
    \item The feature should provide insights into the likelihood of degenerate tensors appearing at or close to a given location.
\end{enumerate}
The features and their use of the ensemble mode definition are as follows:

\subsection*{Ensemble \textit{meanLine}} 
To justify (i), we propose the extraction of the degenerate tensor line of the mean tensor field as a representative, which we refer to as \textit{meanLine}.
This is simply achieved by treating it as a single certain field and extracting degenerate tensor locations as usual.
Note, that a similar definition using the mean mode is not effective, due to the range of the mode between 0 and 1, resulting in a mean value that can only become 1 if all members share the same mode value, thus likely resulting in no feature lines at all.

The advantage of the meanLine - besides its direct link to standard degenerate feature lines - is that its representation can be used to encode additional quantities describing uncertainties, thus justifying (ii).
We call this the \textit{enhanced} meanLine and propose mapping scalar quantities to both the color of the line and the radius of the tube representing the feature line.
In the context of uncertainty quantification, we suggest using the standard deviation of the mode values of all ensemble members at the current line location.
This combines both mode descriptions of the ensemble in a single, yet comprehensive feature indicating the uncertainty of the feature location.
However, a variety of other values such as the mean or standard deviations of fractional anisotropy values could be used.
One could also place an uncertainty tensor glyphs along sampled locations on the meanLine, such as indicated in \cref{fig:componentNoise} (right), which could provide another local description of tensor uncertainty while possibly showing the full tensor information.

\subsection*{Ensemble \textit{modeTube}} 
\begin{figure}[ht!]
    \centering
    \includegraphics[width=\linewidth]{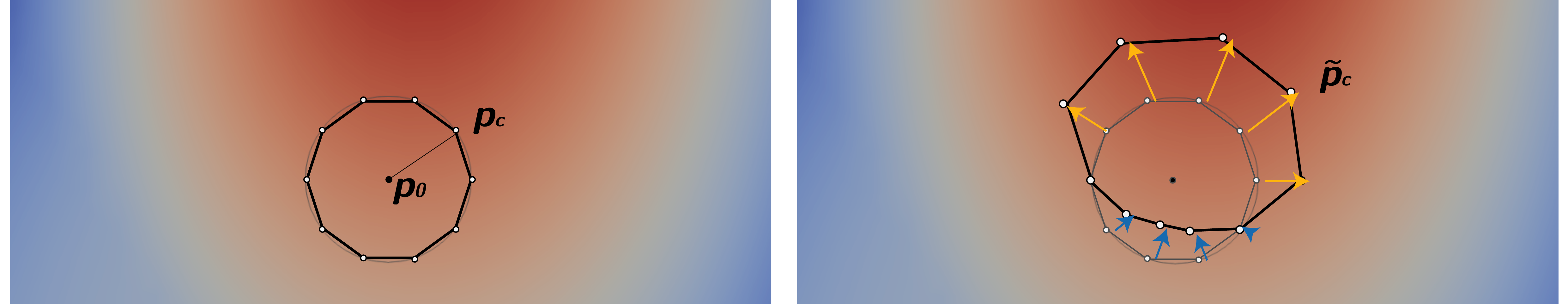}
    \caption{Construction of the Ensemble modeTube: Based on the differences between mean mode value at $\vp_0$ on the meanLine and samples $\vp_c$ on a circle perpendicular to the line tangent, the point locations are moved towards (mean mode at $\vp_0$ is higher) or away from (mean mode at $\vp_0$ is lower) the center, indicating the spatial distribution of mean $\mode_{abs}$ values around the meanLine.
    Mode values are indicated by a divergent cool-to-warm color map.}
    \label{fig:modeTube}
\end{figure}
The (enhanced) meanLine only provides insights into exact locations within the ensemble, due to its sampling on the line location.
However, in ensembles, degenerate tensor locations typically vary from member to member, if existing at all.
Further, as shown in \cref{fig:modeCompare}, mean mode values and the mode of the mean tensor might differ significantly, leading to the mean mode indicating other areas to be more likely to contain degenerate tensors.
To encode the spatial distribution of the mean mode surrounding the meanLine, thus forming a link between both mode descriptions, we propose the construction of the \textit{modeTube}.
For every point of the meanLine $\vp_0$, mean mode is sampled for the point as well as for evenly distributed locations $\vp_c$ on a circle with radius $r_0$ perpendicular to the line tangent.
The resulting plane and sample points are indicated in \cref{fig:modeTube}.
The relation of mean mode at the meanLine to its surrounding mean mode values is given by the distance $d_c = \overline{\mode}(\vp_c) - \overline{\mode}(\vp_0)$, which is then mapped to a displacement vector applied to the sample locations, such that
\begin{equation}\label{eq:modeTube}
    \Tilde{\vp_c} = \vp_0 +  (\vp_c - \vp_0) \cdot f_c \cdot \frac{r_s}{r_0}
\end{equation}
where
\begin{equation}
f_c = \frac{2}{1 + e^{(-2d_c)}}
\end{equation}
and $r_s$ can be used as a scaling factor.
The resulting shape is constructed from the set of points $\Tilde{\vp_c}$ and, therefore, acts similar to a polar plot indicating the directions of higher mean mode locations based on the chosen radius, which provides an alternative solution to (ii).
Additionally, color encoding can be used to indicate the displacement for the sample points, e.g., by applying a divergent color map based on $d_c$.
The two colors thus indicate whether the mean mode value at the center line location is higher or lower than in the direction of the tube sample point.
The combination of both shape and color, therefore, allows to examine the global behavior of the mean mode values surrounding the meanLine.
To account for small differences in values, $d_c$ might be normalized, e.g., by using a global factor such as the largest absolute difference.
One could also consider a normalization per meanLine sample location, which would put the focus on a local encoding of the distribution.
Note, however, using this configuration, a visual comparison of tube locations would only allow one to draw conclusions about the variation of spatial distribution, not of mean mode differences.
In the remainder of this work, we make use of a global normalization factor.

\subsection*{Ensemble \textit{probabilityBand}} 
The features introduced up to this point, even if incorporating the mean mode field, are based on the mean tensor field and the extracted meanLine locations.
This, however, has a clear limitation:
The mean tensor can smooth out critical variations within the tensor fields, i.e., there might not be a meanLine in several areas within the ensemble that are likely to encompass degenerate tensors.
To address this limitation, we propose the \textit{probabilityBand}, a feature surface extracted from uncertain mode values using standard statistical tools.
Consider the distribution of mode values of the ensemble members at every location assuming a Gaussian distribution as in \cref{eq:gauss} with mean mode as in \cref{meanMode} and sample standard deviation $\sigma_{\mode}$.
We are interested in locations, where the probability of a mode distribution to contain values close to 1 is high.
Thus, using a threshold $t$, the probability of the standard normal random variable $X$ at a given ensemble location being equal to or higher than $t$ is 
\begin{equation}
P(X \geq t) = 1 - \Phi(z)
\end{equation}
while the cumulative distribution function $\Phi(z)$ of the standard normal distribution is approximated using the error function such that
\begin{equation}
\Phi(z) = \frac{1}{2} \left[1 + \text{erf}\left(\frac{z}{\sqrt{2}}\right)\right]
\end{equation}
and $z$ is the z-score
\begin{equation}
    z = \frac{t - \overline{\mode}}{\sigma_{\mode}}.
\end{equation}
We compute a scalar field $f(\vx) = P(X \geq t)_{\vx}$ describing these probabilities for each location $\vx$.
When choosing $t$ close to 1, high values indicate a higher likelihood of high mode values, and thus the occurrence of degenerate tensors.
The probabilityBand is then defined as the isosurface $S = \{ \vx \in \RRSet^3 \ | \ f(\vx) = c\}$ extracted from the field using a desired isovalue $c$.
A probabilityBand, therefore, encompasses areas within the ensemble, in which the probability of a mode value $\geq t$ is at $c$ or higher.

\section{Results}\label{sec:results}
To evaluate the effectiveness and properties of the proposed features, we developed a prototype application using C++.
Synthetic and simulated data was provided as VTK unstructured grid data comprised of tetrahedral meshes.
Reading and writing files were handled by the Visualization Toolkit (VTK)~\cite{schroeder2004visualization} in version 9.3. and the extraction of iso-surfaces was done using ParaView~\cite{ParaView} 5.12 and geometry rendered in Blender~\cite{blender} 4.1.
The extraction of degenerate tensor lines was achieved using the implementation of Oster et al.~\cite{oster2018core}, which is available as open-source code on github~\footnote{\url{https://github.com/timo-oster/tensor-lines}}.
The resulting line segments were combined into polylines when sharing common start or end points to allow for an approximation of line tangents using central differences whenever possible. 
Vector and matrix processing was done using Eigen~\footnote{\url{https://eigen.tuxfamily.org/}} in version 3.4.

\subsection{Synthetic data set}
To evaluate our techniques using comprehensible, yet meaningful examples, we created ensembles of synthetic linear tensor fields using the description in \cref{eq:linField}, where the field, and thus, the locations of degenerate tensors can be steered by a couple of parameters.
We choose 
\begin{align}
\begin{split}
    &\mathbf{T}_x =
    \begin{pmatrix}
        1 & 1 & 0\\
        1 & 0 & 0\\
        0 & 0 & 0
    \end{pmatrix}, 
    \quad \mathbf{T}_y =
    \begin{pmatrix}
        0 & -1 & 0\\
        -1 & 2 & 0\\
        0 & 0 & 0
    \end{pmatrix},
    \\&\mathbf{T}_z =  
    \begin{pmatrix}
        0 & 0 & 0\\
        0 & 0 & 0\\
        0 & 0 & 1
    \end{pmatrix},
    \quad \mathbf{T}_0 =  
    \begin{pmatrix}
        7 & 0 & 0\\
        0 & 6 & 0\\
        0 & 0 & 1
    \end{pmatrix}
\end{split} 
\end{align}
with an orthonormal base $x,y,z \in [0,2]$.
This ensures that the tensor eigenvalues are always distinct, 
unless $ x=1 \land y=1$, 
which results in a straight degenerate tensor line along the z-axis at the center of the horizontal plane.
The tensor field can be translated in arbitrary direction $(\Delta x, \Delta y,\Delta z)$ with 
\begin{align}
    \mathbf{T}_{trans}(x,y,z,\Delta x ,\Delta y, \Delta z) = \mathbf{T} (x - \Delta x, y - \Delta y, z - \Delta z).
\end{align}
Further, it can be rotated around the degenerate tensor line by the angle parameter $\theta$ 
\begin{equation}
    \mathbf{T}_{rot} (x,y,z,\theta,\Delta x,\Delta y,\Delta z) = \mathbf{T} (\Tilde{x},\Tilde{y},\Tilde{z}),
\end{equation}
with
\begin{align}\label{affineTransformation}
    \begin{split}
        \begin{pmatrix}
            \Tilde{x}\\
            \Tilde{y}\\
            \Tilde{z}\\
            1
        \end{pmatrix}=
        & \begin{pmatrix}
            1 & 0 & 0 & 1 + \Delta x \\
            0 & 1 & 0 & 1 + \Delta y \\
            0 & 0 & 1 & 0\\
            0 & 0 & 0 & 1
        \end{pmatrix}
        \cdot
        \begin{pmatrix}
            \cos{\theta}& \sin{\theta} & 0 & 0\\
            -\sin{\theta}& \cos{\theta} & 0 & 0\\
            0 & 0 & 1 & 0\\
            0 & 0 & 0 & 1
        \end{pmatrix}\\ &
        \cdot
        \begin{pmatrix}
            1 & 0 & 0 & - 1 - \Delta x \\
            0 & 1 & 0 & - 1 - \Delta y \\
            0 & 0 & 1 & 0\\
            0 & 0 & 0 & 1
        \end{pmatrix}
        \cdot 
        \begin{pmatrix}
            x\\y\\z\\1
        \end{pmatrix}.
    \end{split}
\end{align}
In this manner, we can systematically create ensemble members with desired properties such as specific degenerate tensor line distributions.
It further allows us to examine the impact of the underlying tensor field relative to the actual extracted degenerate tensor lines. 
Two synthetic test ensembles are discussed in the following.
For all members, the domain is uniformly sampled, resulting in a tetrahedron mesh comprising 625,000 cells and 132,651 points each.

\paragraph*{Translated and Rotated Tensor Ensemble}
\begin{figure}[ht!]
    \centering
    \includegraphics[width=\linewidth]{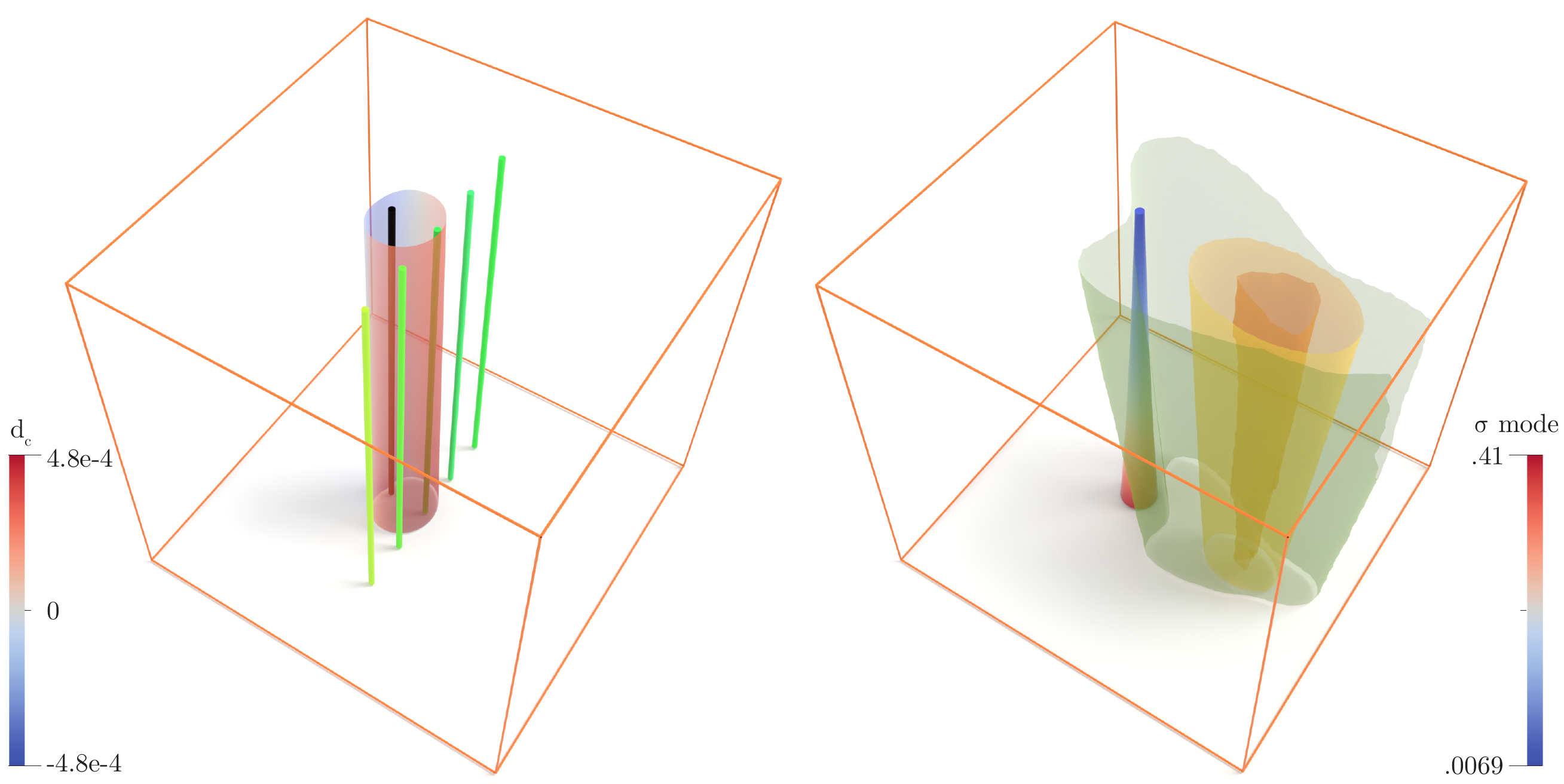}
    \caption{Ensemble comprising five members using a synthetic tensor field. Left: Spaghetti plot showing the extracted degenerate tensor lines of all members each represented by a shade of green and the modeTube indicating the distribution of mean mode values around the meanLine. Color and shape both indicate the difference in mean mode values between the sample locations and the meanline.  Right: Enhanced meanLine indicating degenerate tensor locations of the mean tensor field and the standard deviation of mode values encoded by color and radius. Further, the three nested iso contours indicate different probabilities ($15\%$ \tikzcircle[mylight, fill=mylight]{2pt}, $50\%$ \tikzcircle[mymid, fill=mymid]{2pt}, $90\%$ \tikzcircle[mydark, fill=mydark]{2pt}) of mode values $\geq0.95$ closing in.}
    \label{fig:rotAndTrans}
\end{figure}
The first ensemble consists of five members that are translated along the x-axis with a range of $\Delta x \in [-0.5,0.5]$ and simultaneously rotated around their degenerate tensor lines with a rotation angle range of $\theta \in \left[0,\frac{\pi}{2}\right]$ as indicated by the mode field visualization in \cref{fig:modeCompare}.\\

The five degenerate tensor lines of the members and the corresponding meanLine are illustrated in \cref{fig:rotAndTrans}~(left).
It can be observed that the meanLine is shifted along the y-axis compared to the expected location in the center.
The reason lies in the unique ensemble construction combining both translation and rotation.
However, the proposed uncertainty features can draw attention to such a scenario:
The modeTube forming around the meanLine displayed in \cref{fig:rotAndTrans}~(left) indicates the described shift through form and color, as it moves and turns red towards the center.
Meanwhile, the probabilityBands in \cref{fig:rotAndTrans}~(right) with $t=0.95$ and $c\in [0.15,0.5,0.9] $ show the discrepancy between meanLine and most probable degenerate locations, by not enclosing the meanLine for probabilities $\geq15\%$. 
Yet, choosing a low value for $c$ produces a surface shape that indicates the original distribution of degenerate tensor lines and the mean mode values.

\paragraph*{Component Noise Ensemble}
\begin{figure}[ht!]
    \centering
    \includegraphics[width=0.99\linewidth]{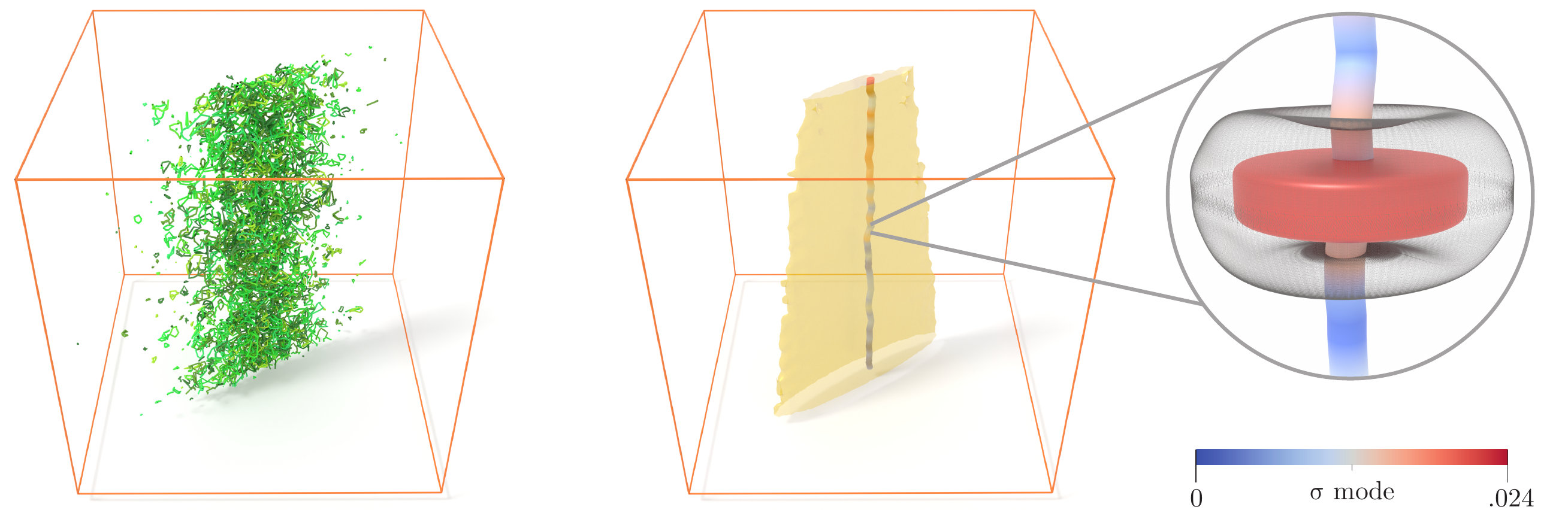}
    \caption{Ensemble comprising 1000 members each constructed by applying random noise to the components of a synthetic tensor field. Left: Spaghetti plot showing the extracted degenerate tensor lines of 10 randomly selected members. Center: Enhanced meanLine with mode standard deviation encoded as color and probabilityBand encompassing locations, where the probability of finding absolute mode values $\geq0.95$ is $33\%$. Right: Uncertain Tensor visualization~\cite{gerrits2019towards} of tensors located on the meanLine using superquadric tensor glyphs~\cite{schultz2010superquadric}.}
    \label{fig:componentNoise}
\end{figure}
The second ensemble contains $1000$ members, each based on the initial tensor field.
However, Gaussian noise $\sim\mathcal{N} (\mu=0,\sigma^2=0.01)$ is added to the individual tensor components sampled for every tensor location.
This results in noisy, unpredictable feature locations for each member, indicated by the spaghetti plot visualization of 10 randomly selected members in \cref{fig:componentNoise}~(left).
Yet, this allows us to both, check if our features can effectively find a representative degenerate line feature close to the original line, and effectively capture the uncertainty provided by the noise.
\\

In contrast to the visual clutter, the corresponding meanLine forms a smooth line in the center of the horizontal plane in \cref{fig:componentNoise}~(right), where in the original field, the degenerate line was placed.
Further, the probabilityBand in \cref{fig:componentNoise}~(right) with $t=0.95$ and $c=0.33$ encloses the area where most of the noisy features are generated.
In combination, the meanLine and the probabilityBand provide a visually simple encoding of the uncertainty.
The spaghetti plot visualization, on the other hand, suffers from severe visual clutter and provides no means of extracting an accurate representative line.
This outcome also suggests that the meanLine could be used as a denoising tool.
Now that a representative feature exists, the optional placement of an uncertain tensor glyph~\cite{gerrits2019towards} allows us to analyze shape variations of the tensors at the meanLine location, introducing another level of detail.

\subsection{Simulation of Stresses in an O-ring}
To test the utility of our new features in the context of mechanical engineering, we extract them from the resulting fields of a simulation describing various stress situations within a mechanical object.
Similar simulations are used to analyze structural deformations, which appear in a large variety of applications.
Topological features not only provide a better overall understanding of the tensor data, but they can also indicate (un-)desired distributions of forces within a material, e.g., through shear deformations.
An analysis of such features in ensembles, therefore, could allow for the identification of trends or areas of interest.\\

Due to its application in previous research, we use different variations of the O-ring data set as used and described by Hung et al.~\cite{hung2023global}.
The simulation contains three parameters describing the periodicity $(p,q)$ and the magnitude of anisotropy $(\alpha)$ of the compression forces.
Conveniently, this allows to control the tensor field and thus the occurrence and location of degenerate features to a degree.
We create three ensembles by selecting simulations with different modified parameters.
Each member comprises 1,953,720 cells and 344,655 points.

\paragraph*{$p$ Variation Ensemble}
\begin{figure*}
    \centering
    \includegraphics[width=\linewidth]{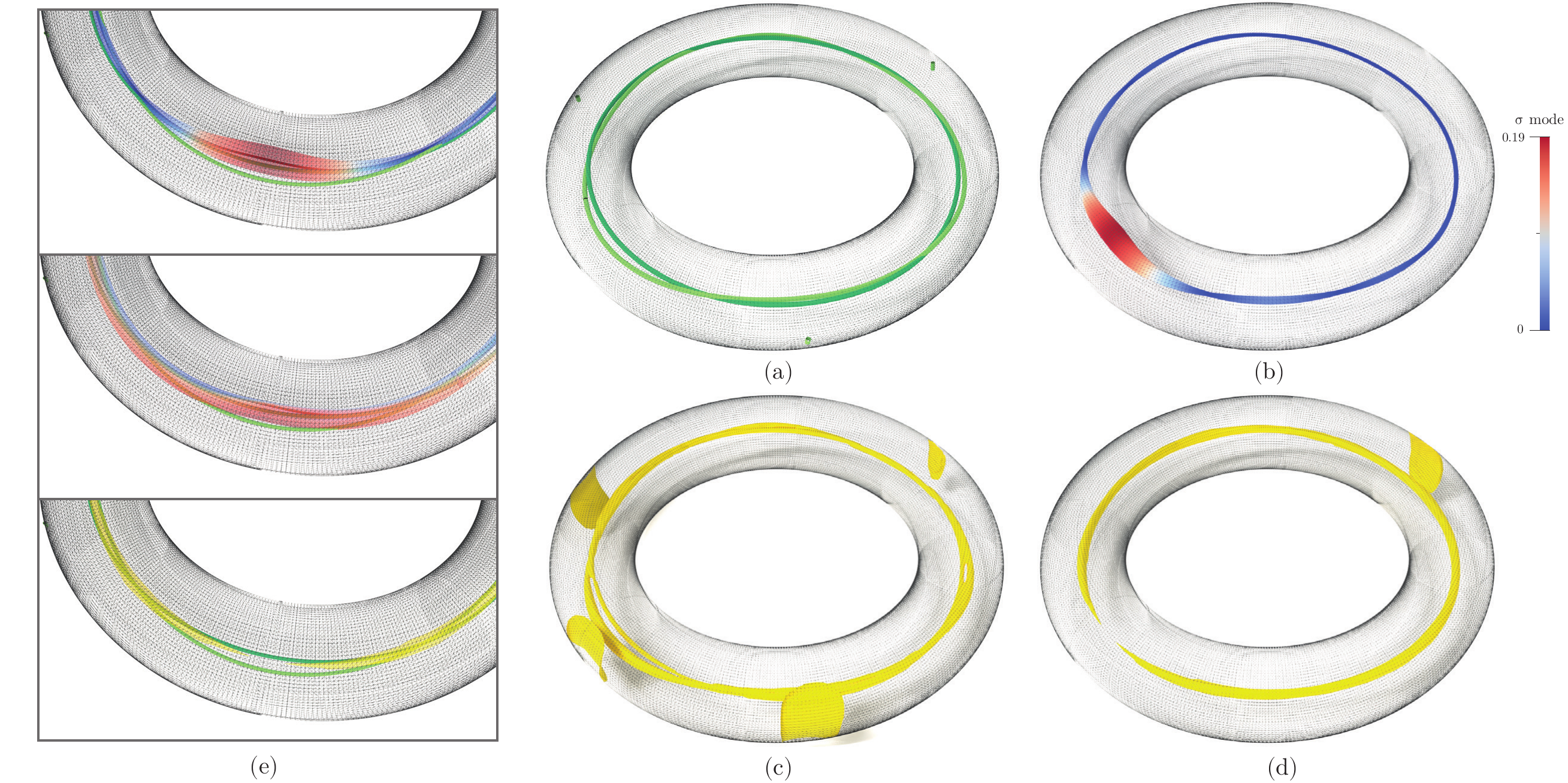}
    \caption{Uncertainty Visualizations for two simulation results describing stresses in an O-ring with varying $p$ parameter. (a) Each member has exactly one degenerate tensor line shown in a different shade of green. (b) The enhanced meanLine shows the location of the degenerate tensor line within the mean tensor field. Radius and color indicate the standard deviation of mode values. (c) probabilityBand indicating locations where mode values have a probability of $50\%$ of a mode value $\geq 0.95$. (d) probabilityBand indicating locations where mode values have a probability of $10\%$ of a mode value $\geq 0.99$. (e) Closeup views show, from top to bottom, an enhanced meanLine, the modeTube, and the probabilityBand visualizations overlaid to the spaghetti plot of degenerate tensor lines of all ensemble members.}
    \label{fig:oringtwo}
\end{figure*}
The first ensemble consists of two members with $q=1$ and $\alpha =35$, differing in parameter $p \in [2,3]$.
The degenerate tensor lines of these members are visualized in a spaghetti plot \cref{fig:oringtwo}~(a) and show that the degenerate tensor lines of both members are almost identical except for a small part in the lower left.
\cref{fig:oringtwo}~(b) shows the resulting enhanced meanLine, where the sampled standard variation of mode values is mapped to color and radius, as well as two probabilityBand visualization with $t = 0.95$, $c = 0.5$ in (c) and $t = 0.99$, $c = 0.1$ in (d).
\cref{fig:oringtwo}~(e) provides a closeup comparison of all features overlaid to the degenerate tensor lines of the ensemble members.\\

When visually comparing the various views, we can see that the behavior indicated by the spaghetti plot is effectively captured by the enhanced meanLine representation:
The meanLine occupies almost the same spatial locations as the extracted degenerate tensor lines except the area, where both lines diverge.
This is additionally encoded by the high mode standard deviation shown in (b) but is also indicated by the modeTube visible in the center closeup, thus indicating the meanLine feature to be less certain at these locations.
Similarly, the probabilityBand indicates a high probability of mode values surrounding the locations of the degenerate tensor lines.
The lower value for $t$ in (c) provides information on locations with high mode values, that are neither represented by the spaghetti plot nor the meanLine and could lead to the appearance of degenerate features in similar simulations.
These features disappear when increasing $t$ as shown in \cref{fig:oringtwo}~(d) and also lead to a visual gap at the location of diverging lines, further supporting our assumption of assuming a lower likelihood of finding degenerate tensor lines at these locations.

\paragraph*{$\alpha$ Variation Ensemble}
\begin{figure*}
    \centering
    \includegraphics[width=\linewidth]{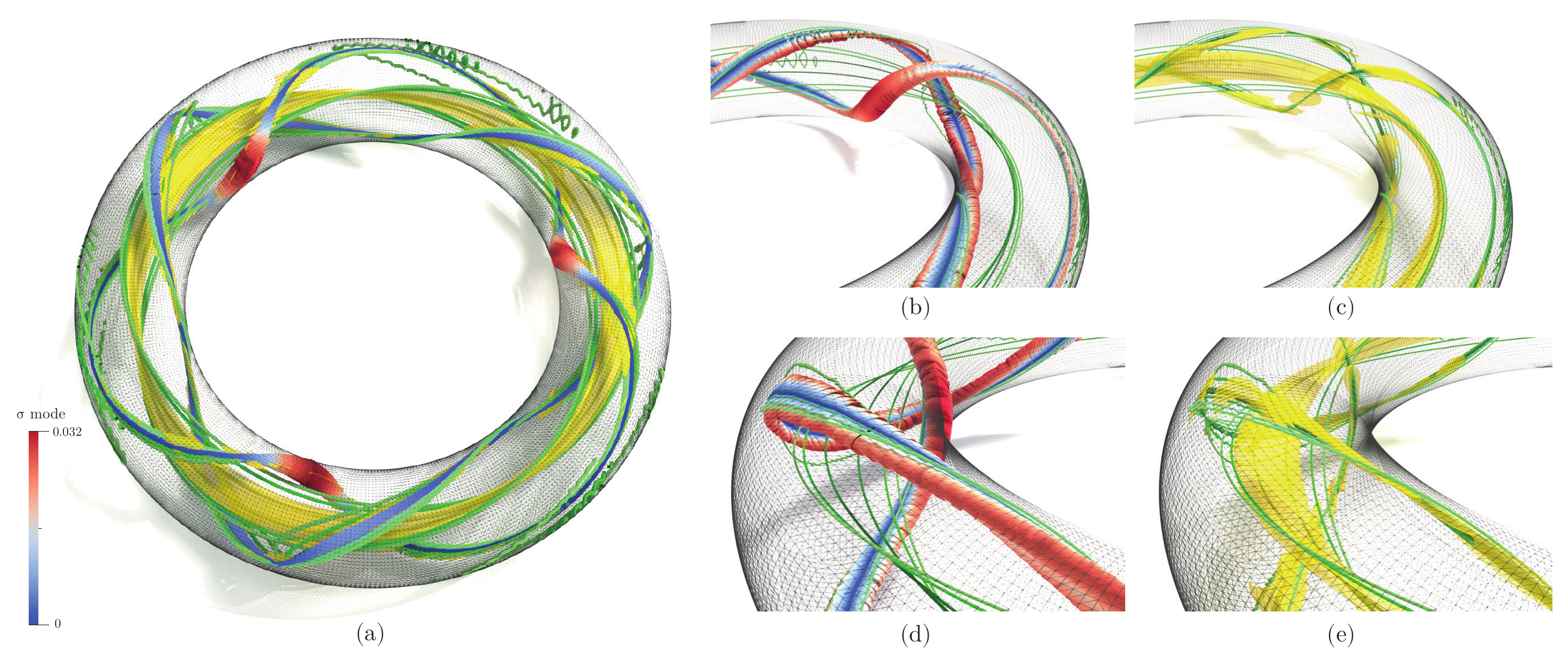}
    \caption{Uncertainty Visualizations for eight simulation results describing stresses in an O-ring with varying $\alpha$ parameter. (a) Spaghetti plot of degenerate tensor lines of all ensemble members, each represented by a different shade of green as well as an enhanced meanLine showing the locations of degenerate tensors within the mean tensor field. Color and radius encode standard deviation of the mode values. The probabilityBand in yellow indicates locations where mode values have a probability of $25\%$ of a mode value $\geq 0.99$. (b) and (d) Closeup view showing how modeTube indicates the spread of extracted degenerate tensor lines close to the meanLines. Both color and shape variation show the difference between mean mode values at the tube sample location and the meanLine.  (c) and (e) Closeup view showing how the probabilityBand indicates the spread of extracted degenerate tensor lines in the whole field.}
    \label{fig:oringavar}
\end{figure*}
The second ensemble contains eight members with $p = 3$, $q = 2$, and $\alpha \in [20, 25, 30, 33, 34, 35, 40, 45]$ and therefore describes a variation of the anisotropy magnitude.
The degenerate tensor lines extracted from the separate members are shown in \cref{fig:teaser} (left) and overlaid with different feature variations in \cref{fig:oringavar}.\\

Again, we start our analysis from the spaghetti plot visualization and evaluate, if our novel features provide similar or additional insights.
A first visual analysis indicates two groups of lines that could be clustered together, as their occupied space through the domain is reminiscent of twisting ribbons, whereas one is less broad than the other, as lines stay closer to each other.
When looking at \cref{fig:oringavar}~(a), it becomes evident that both groups are captured by our proposed features.
In this case, the narrow ribbon is represented by the enhanced meanLine, which runs along their center, while the broader one is captured by the probabilityBand.
The closeup renderings in \cref{fig:oringavar}~(c) and (e) show a continuous surface that follows the twisting motion of the lines.
However, some smaller and fragmented surfaces can also be found in the other group.
While the enhanced meanLine shown in \cref{fig:teaser} (center) and \cref{fig:oringavar}~(a), provides some indication of meanLine uncertainty through the coloring and radius, the modeTube is capable of capturing the spatial spread of the lines surrounding the features, as shown in \cref{fig:oringavar}~(b) and (e).
Thus, while the spaghetti plot visualization might help in cases where they don't produce visual clutter, the rendering in \cref{fig:teaser} (center and left) suggests, that our proposed features offer a similar, if not more intricate insight into the ensemble.

\paragraph*{$q$ Variation Ensemble}
\begin{figure}[ht!]
    \centering
    \includegraphics[width=\linewidth]{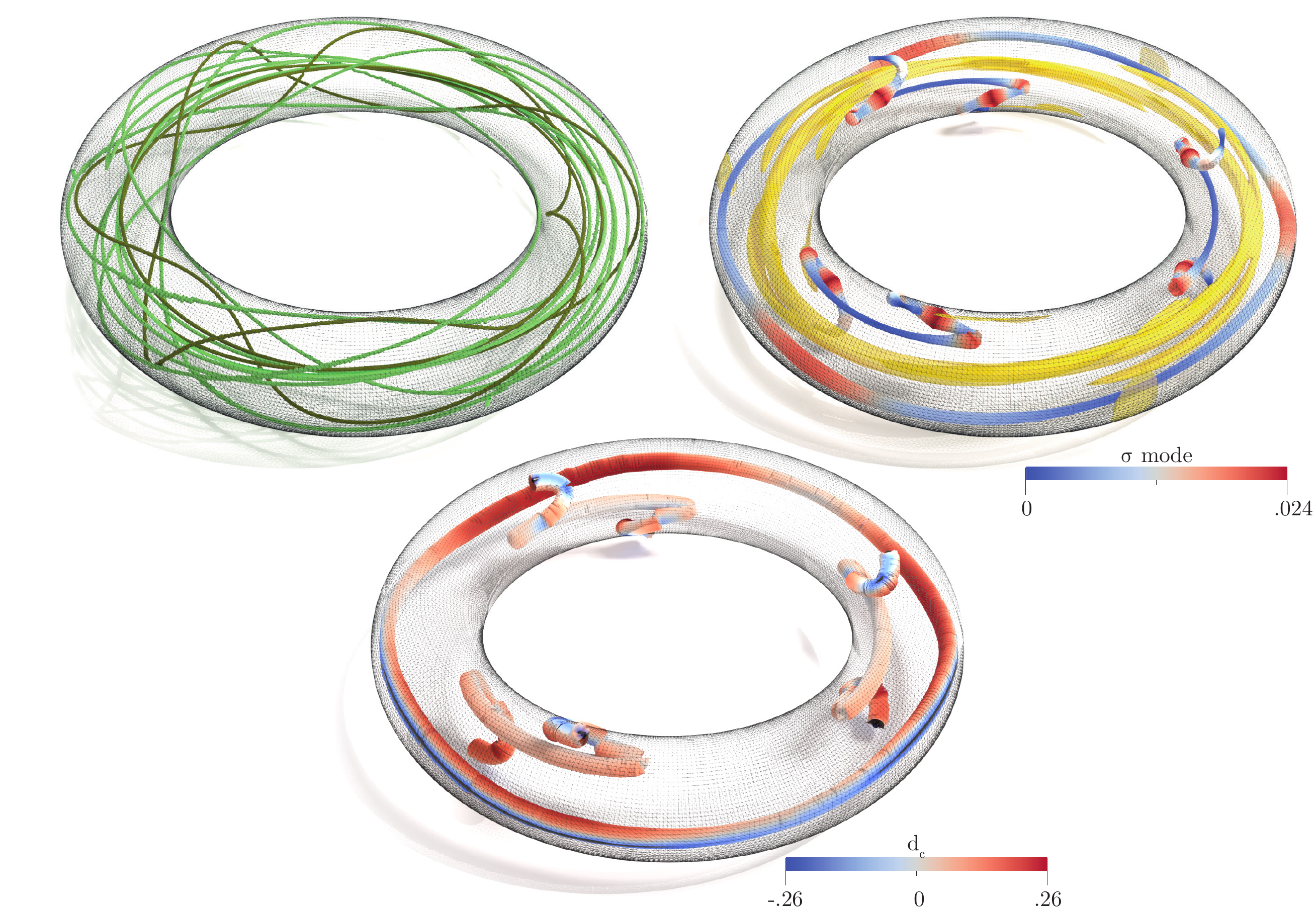}
    \caption{Uncertainty visualizations for three simulation results describing stresses in an O-ring with varying $q$ parameter. Left: Spaghetti plot of degenerate tensor lines of all ensemble members, each represented by a different shade of green. Right: Enhanced meanLine showing the location of the degenerate tensors within the mean tensor field. Color and radius encode standard deviation of the mode values. Further, the yellow probabilityBand indicates locations where mode values have a probability of $20\%$ of a mode value $\geq 0.99$.
    Center: modeTube visualization where shape and color indicate spatial distribution and differences of mean mode value in the vicinity of the meanLine.}
    \label{fig:oringqvar}
\end{figure}
For the last ensemble $p=3$ and $\alpha=35$ are kept constant, while $q \in [1,2,3]$.
The resulting ensemble of three members is portrayed in \cref{fig:oringqvar} through a spaghetti plot (left), as well as the enhanced meanLine and a probabilityBand with $t = 0.99$ and $c = 0.2$ (right) and the modeTube (center).\\

Compared to the previous collections, the extracted degenerate tensor lines differ significantly and no visually clear pattern emerges.
When looking closely, one ribbon-like cluster in the center of the ring can be found.
Extracting the meanLine results in four curves, one showing the typical loop behavior of the other lines and three that show a back-and-forth motion that connects the top side of the O-ring with the bottom side.
Due to the variability in the underlying tensors, even the enhanced meanLine does not fully capture the behavior that we extract from looking at the spaghetti plot.
However, the modeTube shape and color indicate directions of higher mode values towards the center of the O-ring.
These align with the location of extracted degenerate tensor lines in the ensemble.
Here, again, the aforementioned ribbon is captured by the probabilityBand visualization (right).

\section{Discussion}
In the following, we want to discuss choices made in the design of our proposed features as well as possible alternatives.
First of all, we base the description of degenerate tensor locations on the tensor mode.
There exist alternative tensor descriptions that relate to eigenvalue relations, such as anisotropy measures~\cite{tricoche2008invariant}, or the tensor discriminant~\cite{zheng2005topological}.
The mode, however, has developed into the de facto standard quantity for the extraction of degenerate tensor locations and its interpretation is straightforward.
The symmetry of the mode value further allows us to make the simplification of using $\mode_{abs}$ to indicate degenerate tensor locations.
While this could also be viewed as a limitation of our approach, in this work, our focus lies on describing the uncertainty of double-degenerate tensor locations regardless of their classification.
A post-hoc analysis of the distribution of linear/planar degeneracies at the feature locations might be a way to address this, but this endeavor is planned for future work.\\

When comparing the different features introduced, one might not see the necessity for all of them.
For some examples, there actually appear to be redundancies communicated by the different visualizations.
However, we believe, that each feature has a unique focus, that cannot be fully covered by the respective others.
For most cases we tested, the (enhanced) meanLine provided a clear, parameter-free  representation of degenerate tensor line behavior within the ensemble.
Especially lines close to each other were almost always captured by a meanLine in the mean tensor field and the modeTube provided information on the local spread in the vicinity of the meanLine.
As the meanLine is based on the definition of the degenerate tensor line of a certain field and is thus a line-type feature, it suffers from the same problematic cases, such as the formation of degenerate planes within the data as demonstrated in \cite{oster2018parallel}.
However, these are structurally unstable and rarely appear in non-constructed scenarios.
Only when members vary a lot, no meanLine might be present, such as in \cref{fig:oringavar}~(b) and (d) or it might show some unexpected behavior as seen in \cref{fig:oringqvar}.
While the modeTube might indicate the spatial distribution of these missing features, however, to truly compensate for this restriction, the probabilityBand is capable of capturing those areas.
The bands are, however, less straightforward to interpret and depend on the choice of parameters, which we will discuss later.
They can easily comprise fragmented surfaces and complex nested structures as seen in \cref{fig:oringqvarprob}.
For all cases presented in \cref{sec:results}, the features complement each other thus providing a more comprehensive description of the ensemble.
As (enhanced) meanLine and probabilityBand are based on two different descriptions of the mode distribution within the ensemble, we see this as a further hint supporting our assumption, that both are needed to describe degenerate tensor uncertainty comprehensively.\\

While in this work, we use spaghetti plots as reference visualizations to indicate degenerate tensor behavior of the members within our selected test data sets, they do not represent a valuable option for representing uncertainties of a whole ensemble in the general case.
As briefly mentioned in the background section, spaghetti plots fail to account for the fact that some members might have line-type features, while others do not.
Further, they can easily lead to visual clutter and become computationally infeasible for large ensembles, which is already indicated in \cref{fig:componentNoise} (left), where the features of only 10 of all 1000 members already produce a cluttered view.

\subsection*{Parameters and Models}
\begin{figure}[ht!]
    \centering
    \includegraphics[width=\linewidth]{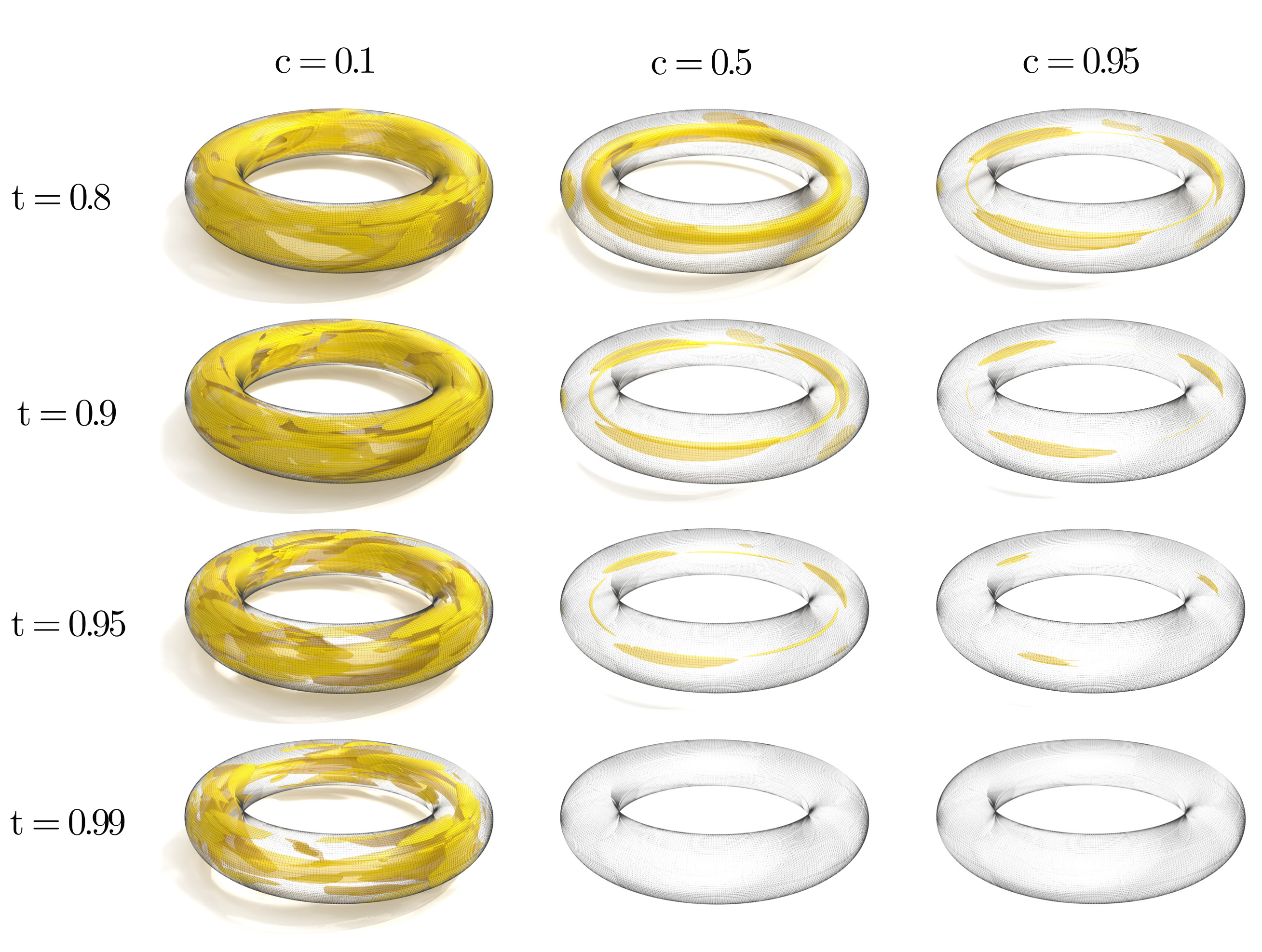}
    \caption{Impact of parameters on the probabilityBand in the $q$ variation ensemble of stresses in an O-ring. Scalar fields describing the probability of a location to inhibit a mode value $\geq t$ are used for the extraction of isosurfaces with isovalue $c$ representing $10\%$, $50\%$, and $95\%$ probability respectively.
    }
    \label{fig:oringqvarprob}
\end{figure}
Two of our proposed features rely on choosing suitable parameters.
The first is the radius $r_0$ used in the construction of the modeTube in \cref{eq:modeTube}.
Modifying it changes the values compared to the center point, resulting in different modeTube shapes.
Even when decoupling the sampling locations from the initial circle geometry positions, e.g., by using a suitable scaling factor $r_s$ or introducing a minimum radius, a value too small might lead to almost imperceptible directional variations, while a radius too large might miss important changes in the field.
For all results shown in this work, we relied on a radius of $r_0 = 0.01$, based on empirical tests and the dimensions of test datasets.
Further, deriving an accurate understanding of the distribution of mean mode values based on the modeTube shape necessitates knowing either the original sampling- or the meanLine locations.
Adding these to the visualization, e.g., by rendering the meanLine within a transparent modeTube can lead to visual clutter and other visual artifacts.
While the use of color as proposed in this work can provide some help, alternatives, such as interactively changing views or transparency, or finding alternative mappings that include a reference to the original center of the tube could be explored.\\

The probabilityBand introduces two parameters, namely a threshold $t$, and the isovalue $c$, which can be used to steer the resulting surface.
The impact of changing parameters is less straightforward, which can be seen in the results of a small parameter study shown in \cref{fig:oringqvarprob}.
Based on our observations, a value of $t$ chosen closer to $1$ can be interpreted as putting a stronger focus on tensors actually becoming degenerate at a certain location while choosing the isovalue allows us to generate information on the distribution of values at this location and thus on the confidence of the occurrence of a feature.
As can be seen in \cref{fig:oringqvarprob} for results where $t = 0.99$, setting $c = 0.5$ and $c = 0.95$ do not produce a probabilityBand, as no location within the field shows such a significant probability of mode values exceeding the high mode value. 
Thus, one needs to adjust both values based on the analysis question.
If interested in locations that are most likely to form degenerate tensor lines in the ensemble members, $t$ should be chosen high.
If, on the other hand, the goal is to gain a general understanding of the distribution of mode values within the ensemble, a lower $t$ can be used and the probability analyzed using the isovalue $c$.
Alternatively, volume rendering approaches could be considered to see all probabilities for a given $t$, but this might result in visually more challenging representations.\\

In our work, we assume that the members of the tensor ensembles follow a single Gaussian distribution and further apply the same reasoning to the tensor mode values of ensemble members.
This assumption was not only similarly applied and justified in related works (e.g., ~\cite{gerrits2019towards, basser2003normal, pothkow2012uncertainty}), but it further significantly simplifies the definition and calculation of the presented features.
However, there are cases where these assumptions do not hold, especially for smaller ensembles, which might require more advanced descriptions of the distribution, or where features would be better captured by two independent distributions.
Further, the absolute mode values used here can only take values in the range $[0,1]$.
Assuming normally distributed values can take values outside of this range.
Using an alternative, such as a truncated normal distribution or a beta distribution might, therefore, result in a more accurate description of probabilities.
Similarly, we use a simple component-wise mean as the tensor mean, which might, as mentioned before, not necessarily interpolate tensor invariants similarly~\cite{gahm2012linear, hotz2010tensor}.
While we addressed this by using the distribution of tensor modes for the probabilityBand, one might consider using a different definition of the mean tensor.
In preliminary tests, however, these often did not result in feature lines within the mean tensor field, e.g., a shape-preserving tensor averaging method would only produce degenerate features at locations where all members hold a degenerate tensor.
This issue needs to be investigated further.

\subsection*{Performance}
While not a particular focus of this work, note that all presented operations and feature definitions allow for a fast generation, extraction, and visualization, as they are computationally simple and many aspects of the computation can be parallelized, e.g., using graphics hardware.
In our implementation, the largest amount of computation time was used for the extraction of degenerate feature lines, followed by I/O operations.
While a more recent implementation of an extraction method, e.g., by Roy et al.~\cite{roy2018robust} might reduce the computation time, only a single line extraction is necessary for our approaches.
Therefore, our features are suitable to be used for the interactive visual analysis of tensor ensembles, where the effect of parameter changes could be explored in real-time.

\section{Conclusion and Future Work}
In this work, we explored the description and visualization of degenerate tensors within three-dimensional symmetric second-order tensor field ensembles.
The goal was to develop and evaluate effective visual features to represent the overall behavior and uncertainty of degenerate tensors within the ensemble.
Such features currently represent a gap in the literature but could be beneficial for the analysis of tensor data.
We proposed two descriptions of the tensor mode within the ensemble to represent degenerate tensor locations and introduced a number of novel features, each communicating different aspects of the underlying distribution of degenerate tensors.  
Using synthetic datasets and simulation ensembles, we were able to show that these new features can effectively describe the general trend of degenerate tensors within the ensemble while simultaneously providing information on the uncertainty of the features.
The results indicate, that especially the combination of two different descriptions of mode values within an ensemble provides a more comprehensive description of the behavior of the ensemble members in regard to degenerate tensor locations.
To the best of our knowledge, we know of no existing techniques, that provide similar insights into tensor ensembles.\\

We believe that the ideas presented in this work are a first step and can provide a starting point for further developments in the visualization of degenerate tensor locations within symmetric second-order tensor ensembles.
Naturally, besides some aspects already mentioned in the discussion, several properties and design decisions used within this work need to be further examined and tested, while alternative models need to be explored, e.g., using descriptions such as fractional anisotropy.
This includes ensembles comprising even higher numbers of members as well as representing examples from other domains, such as diffusion tensor imaging.
While in preliminary tests, alternative descriptions of the tensor distribution and mean tensor did not lead to satisfactory results, a more thorough mathematical analysis of the used properties and alternatives will be explored in the future.
A user study or expert interviews could further provide insights into the usability and effectiveness of our features, allowing for changes in the design or integration in other analysis workflows.
Finally, for this work, we only considered symmetric second-order tensor fields.
Thus, we plan to look into the requirements to provide similar descriptions for general, asymmetric tensor fields, which currently lack advanced tools to visualize their uncertainty, or higher-order tensor ensembles \cite{schultz2011topological}.

\acknowledgments{
The authors wish to thank Eugene Zhang and Yue Zhang for providing the O-ring dataset. The authors further gratefully acknowledge the German Federal Ministry of Education and Research (BMBF) and the NRW state government for supporting this work/project as part of the NHR funding.}

\bibliographystyle{abbrv-doi}

\bibliography{template}

\begin{thebibliography}{10}

\bibitem{abbasloo2015visualizing}
A.~Abbasloo, V.~Wiens, M.~Hermann, and T.~Schultz.
\newblock Visualizing tensor normal distributions at multiple levels of detail.
\newblock {\em IEEE Transactions on Visualization and Computer Graphics}, 22(1):975–984, jan 2016. doi: {{%
10\hspace{.1pt}\discretionary{.}{%
}{.}\hspace{.4pt}1109\discretionary{/}{%
}{/}TVCG\hspace{.1pt}\discretionary{.}{%
}{.}\hspace{.4pt}2015\hspace{.1pt}\discretionary{.}{%
}{.}\hspace{.4pt}2467031}}


\bibitem{ParaView}
J.~Ahrens, B.~Geveci, and C.~Law.
\newblock Paraview: An end-user tool for large data visualization. the visualization handbook.
\newblock pp. 717--731. Elsevier Inc., 2005.

\bibitem{basser2003normal}
P.~Basser and S.~Pajevic.
\newblock A normal distribution for tensor-valued random variables: applications to diffusion tensor mri.
\newblock {\em IEEE Transactions on Medical Imaging}, 22(7):785--794, 2003. doi: {{%
10\hspace{.1pt}\discretionary{.}{%
}{.}\hspace{.4pt}1109\discretionary{/}{%
}{/}TMI\hspace{.1pt}\discretionary{.}{%
}{.}\hspace{.4pt}2003\hspace{.1pt}\discretionary{.}{%
}{.}\hspace{.4pt}815059}}


\bibitem{basser2007spectral}
P.~J. Basser and S.~Pajevic.
\newblock Spectral decomposition of a 4th-order covariance tensor: Applications to diffusion tensor mri.
\newblock {\em Signal Processing}, 87(2):220--236, 2007. doi: {{%
10\hspace{.1pt}\discretionary{.}{%
}{.}\hspace{.4pt}1016\discretionary{/}{%
}{/}j\hspace{.1pt}\discretionary{.}{%
}{.}\hspace{.4pt}sigpro\hspace{.1pt}\discretionary{.}{%
}{.}\hspace{.4pt}2006\hspace{.1pt}\discretionary{.}{%
}{.}\hspace{.4pt}02\hspace{.1pt}\discretionary{.}{%
}{.}\hspace{.4pt}050}}


\bibitem{bi2019survey}
C.~Bi, L.~Yang, Y.~Duan, and Y.~Shi.
\newblock A survey on visualization of tensor field.
\newblock {\em Journal of Visualization}, 22:641--660, 2019. doi: {{%
10\hspace{.1pt}\discretionary{.}{%
}{.}\hspace{.4pt}1007\discretionary{/}{%
}{/}s12650\discretionary{%
}{-}{-}019\discretionary{%
}{-}{-}00555\discretionary{%
}{-}{-}8}}


\bibitem{bonneau2014overview}
G.-P. Bonneau, H.-C. Hege, C.~R. Johnson, M.~M. Oliveira, K.~Potter, P.~Rheingans, and T.~Schultz.
\newblock Overview and state-of-the-art of uncertainty visualization.
\newblock {\em Scientific Visualization: Uncertainty, Multifield, Biomedical, and Scalable Visualization}, pp. 3--27, 2014. doi: {{%
10\hspace{.1pt}\discretionary{.}{%
}{.}\hspace{.4pt}1007\discretionary{/}{%
}{/}978\discretionary{%
}{-}{-}1\discretionary{%
}{-}{-}4471\discretionary{%
}{-}{-}6497\discretionary{%
}{-}{-}5\_1}}


\bibitem{chaves2024depth}
N.~F. Chaves-de Plaza, M.~Molenaar, P.~Mody, M.~Staring, R.~van Egmond, E.~Eisemann, A.~Vilanova, and K.~Hildebrandt.
\newblock Depth for multi-modal contour ensembles.
\newblock In {\em Computer Graphics Forum}, p. e15083. Wiley Online Library, 2024. doi: {{%
10\hspace{.1pt}\discretionary{.}{%
}{.}\hspace{.4pt}1111\discretionary{/}{%
}{/}cgf\hspace{.1pt}\discretionary{.}{%
}{.}\hspace{.4pt}15083}}


\bibitem{blender}
B.~O. Community.
\newblock {\em Blender - a 3D modelling and rendering package}.
\newblock Blender Foundation, Stichting Blender Foundation, Amsterdam, 2018.

\bibitem{delmarcelle1993visualizing}
T.~Delmarcelle and L.~Hesselink.
\newblock Visualizing second-order tensor fields with hyperstreamlines.
\newblock {\em IEEE Computer Graphics and Applications}, 13(4):25--33, 1993. doi: {{%
10\hspace{.1pt}\discretionary{.}{%
}{.}\hspace{.4pt}1109\discretionary{/}{%
}{/}38\hspace{.1pt}\discretionary{.}{%
}{.}\hspace{.4pt}219447}}


\bibitem{fu2014topologically}
F.~Fu and N.~M. Abukhdeir.
\newblock A topologically-informed hyperstreamline seeding method for alignment tensor fields.
\newblock {\em IEEE Transactions on Visualization and Computer Graphics}, 21(3):413--419, 2015. doi: {{%
10\hspace{.1pt}\discretionary{.}{%
}{.}\hspace{.4pt}1109\discretionary{/}{%
}{/}TVCG\hspace{.1pt}\discretionary{.}{%
}{.}\hspace{.4pt}2014\hspace{.1pt}\discretionary{.}{%
}{.}\hspace{.4pt}2363828}}


\bibitem{gahm2012linear}
J.~K. Gahm, N.~Wisniewski, G.~Kindlmann, G.~L. Kung, W.~S. Klug, A.~Garfinkel, and D.~B. Ennis.
\newblock Linear invariant tensor interpolation applied to cardiac diffusion tensor mri.
\newblock In {\em Medical Image Computing and Computer-Assisted Intervention -- MICCAI 2012}, pp. 494--501. Springer Berlin Heidelberg, 2012. doi: {{%
10\hspace{.1pt}\discretionary{.}{%
}{.}\hspace{.4pt}1007\discretionary{/}{%
}{/}978\discretionary{%
}{-}{-}3\discretionary{%
}{-}{-}642\discretionary{%
}{-}{-}33418\discretionary{%
}{-}{-}4\_61}}


\bibitem{gerrits2016glyphs}
T.~Gerrits, C.~R{\"o}ssl, and H.~Theisel.
\newblock Glyphs for general second-order 2d and 3d tensors.
\newblock In {\em IEEE transactions on visualization and computer graphics}, vol.~23, pp. 980--989. IEEE, 2016. doi: {{%
10\hspace{.1pt}\discretionary{.}{%
}{.}\hspace{.4pt}1109\discretionary{/}{%
}{/}TVCG\hspace{.1pt}\discretionary{.}{%
}{.}\hspace{.4pt}2016\hspace{.1pt}\discretionary{.}{%
}{.}\hspace{.4pt}2598998}}


\bibitem{gerrits2018approximate}
T.~Gerrits, C.~R{\"o}ssl, and H.~Theisel.
\newblock An approximate parallel vectors operator for multiple vector fields.
\newblock In {\em Computer Graphics Forum}, vol.~37, pp. 315--326, 2018. doi: {{%
10\hspace{.1pt}\discretionary{.}{%
}{.}\hspace{.4pt}1111\discretionary{/}{%
}{/}cgf\hspace{.1pt}\discretionary{.}{%
}{.}\hspace{.4pt}13422}}


\bibitem{gerrits2019towards}
T.~Gerrits, C.~R{\"o}ssl, and H.~Theisel.
\newblock Towards glyphs for uncertain symmetric second-order tensors.
\newblock In {\em Computer Graphics Forum}, vol.~38, pp. 325--336. Wiley Online Library, 2019. doi: {{%
10\hspace{.1pt}\discretionary{.}{%
}{.}\hspace{.4pt}1111\discretionary{/}{%
}{/}cgf\hspace{.1pt}\discretionary{.}{%
}{.}\hspace{.4pt}13692}}


\bibitem{hergl2021visualization}
C.~Hergl, C.~Blecha, V.~Kretzschmar, F.~Raith, F.~Günther, M.~Stommel, J.~Jankowai, I.~Hotz, T.~Nagel, and G.~Scheuermann.
\newblock Visualization of tensor fields in mechanics.
\newblock In {\em Computer Graphics Forum}, vol.~40, pp. 135--161, 2021. doi: {{%
10\hspace{.1pt}\discretionary{.}{%
}{.}\hspace{.4pt}1111\discretionary{/}{%
}{/}cgf\hspace{.1pt}\discretionary{.}{%
}{.}\hspace{.4pt}14209}}


\bibitem{hesselink1997topology}
L.~Hesselink, Y.~Levy, and Y.~Lavin.
\newblock The topology of symmetric, second-order 3d tensor fields.
\newblock {\em IEEE Transactions on Visualization and Computer Graphics}, 3(1):1–11, jan 1997. doi: {{%
10\hspace{.1pt}\discretionary{.}{%
}{.}\hspace{.4pt}1109\discretionary{/}{%
}{/}2945\hspace{.1pt}\discretionary{.}{%
}{.}\hspace{.4pt}582332}}


\bibitem{hlawitschka2014top}
M.~Hlawitschka, I.~Hotz, A.~Kratz, G.~E. Marai, R.~Moreno, G.~Scheuermann, M.~Stommel, A.~Wiebel, and E.~Zhang.
\newblock Top challenges in the visualization of engineering tensor fields.
\newblock In {\em Visualization and Processing of Tensors and Higher Order Descriptors for Multi-Valued Data}, pp. 3--15. Springer Berlin Heidelberg, Berlin, Heidelberg, 2014.

\bibitem{hotz2010tensor}
I.~Hotz, J.~Sreevalsan-Nair, and B.~Hamann.
\newblock Tensor field reconstruction based on eigenvector and eigenvalue interpolation.
\newblock {\em Scientific Visualization: Advanced Concepts}, 1:110 -- 123, 2010. doi: {{%
10\hspace{.1pt}\discretionary{.}{%
}{.}\hspace{.4pt}4230\discretionary{/}{%
}{/}DFU\hspace{.1pt}\discretionary{.}{%
}{.}\hspace{.4pt}SciViz\hspace{.1pt}\discretionary{.}{%
}{.}\hspace{.4pt}2010\hspace{.1pt}\discretionary{.}{%
}{.}\hspace{.4pt}110}}


\bibitem{hung2021feature}
S.-H. Hung, Y.~Zhang, H.~Yeh, and E.~Zhang.
\newblock Feature curves and surfaces of 3d asymmetric tensor fields.
\newblock {\em IEEE Transactions on Visualization and Computer Graphics}, 28(1):33–42, jan 2022. doi: {{%
10\hspace{.1pt}\discretionary{.}{%
}{.}\hspace{.4pt}1109\discretionary{/}{%
}{/}TVCG\hspace{.1pt}\discretionary{.}{%
}{.}\hspace{.4pt}2021\hspace{.1pt}\discretionary{.}{%
}{.}\hspace{.4pt}3114808}}


\bibitem{hung2023global}
S.-H. Hung, Y.~Zhang, and E.~Zhang.
\newblock Global topology of 3d symmetric tensor fields.
\newblock {\em IEEE Transactions on Visualization and Computer Graphics}, 30(1):1282--1291, 2024. doi: {{%
10\hspace{.1pt}\discretionary{.}{%
}{.}\hspace{.4pt}1109\discretionary{/}{%
}{/}TVCG\hspace{.1pt}\discretionary{.}{%
}{.}\hspace{.4pt}2023\hspace{.1pt}\discretionary{.}{%
}{.}\hspace{.4pt}3326933}}


\bibitem{isenberg2015survey}
T.~Isenberg.
\newblock A survey of illustrative visualization techniques for diffusion-weighted mri tractography.
\newblock In {\em Visualization and Processing of Higher Order Descriptors for Multi-Valued Data}, pp. 235--256. Springer International Publishing, Cham, 2015.

\bibitem{jiao2012uncertainty}
F.~Jiao, J.~M. Phillips, Y.~Gur, and C.~R. Johnson.
\newblock Uncertainty visualization in hardi based on ensembles of odfs.
\newblock In {\em 2012 IEEE Pacific Visualization Symposium}, pp. 193--200, 2012. doi: {{%
10\hspace{.1pt}\discretionary{.}{%
}{.}\hspace{.4pt}1109\discretionary{/}{%
}{/}PacificVis\hspace{.1pt}\discretionary{.}{%
}{.}\hspace{.4pt}2012\hspace{.1pt}\discretionary{.}{%
}{.}\hspace{.4pt}6183591}}


\bibitem{jiao2010metrics}
F.~Jiao, J.~M. Phillips, J.~Stinstra, J.~Kr\"{u}ger, R.~Varma, E.~Hsu, J.~Korenberg, and C.~R. Johnson.
\newblock Metrics for uncertainty analysis and visualization of diffusion tensor images.
\newblock In {\em Proceedings of the 5th International Conference on Medical Imaging and Augmented Reality}, p. 179–190. Springer-Verlag, Berlin, Heidelberg, 2010.

\bibitem{jones2002spatial}
D.~K. Jones, L.~D. Griffin, D.~C. Alexander, M.~Catani, M.~A. Horsfield, R.~Howard, and S.~C. Williams.
\newblock Spatial normalization and averaging of diffusion tensor mri data sets.
\newblock {\em NeuroImage}, 17(2):592--617, 2002. doi: {{%
10\hspace{.1pt}\discretionary{.}{%
}{.}\hspace{.4pt}1006\discretionary{/}{%
}{/}nimg\hspace{.1pt}\discretionary{.}{%
}{.}\hspace{.4pt}2002\hspace{.1pt}\discretionary{.}{%
}{.}\hspace{.4pt}1148}}


\bibitem{kratz2013visualization}
A.~Kratz, C.~Auer, M.~Stommel, and I.~Hotz.
\newblock Visualization and analysis of second-order tensors: Moving beyond the symmetric positive-definite case.
\newblock In {\em Computer Graphics Forum}, vol.~32, pp. 49--74, 2013. doi: {{%
10\hspace{.1pt}\discretionary{.}{%
}{.}\hspace{.4pt}1111\discretionary{/}{%
}{/}j\hspace{.1pt}\discretionary{.}{%
}{.}\hspace{.4pt}1467\discretionary{%
}{-}{-}8659\hspace{.1pt}\discretionary{.}{%
}{.}\hspace{.4pt}2012\hspace{.1pt}\discretionary{.}{%
}{.}\hspace{.4pt}03231\hspace{.1pt}\discretionary{.}{%
}{.}\hspace{.4pt}x}}


\bibitem{kratz2011visual}
A.~Kratz, B.~Meyer, and I.~Hotz.
\newblock {A Visual Approach to Analysis of Stress Tensor Fields}.
\newblock {\em Scientific Visualization: Interactions, Features, Metaphors}, 2:188--211, 2011. doi: {{%
10\hspace{.1pt}\discretionary{.}{%
}{.}\hspace{.4pt}4230\discretionary{/}{%
}{/}DFU\hspace{.1pt}\discretionary{.}{%
}{.}\hspace{.4pt}Vol2\hspace{.1pt}\discretionary{.}{%
}{.}\hspace{.4pt}SciViz\hspace{.1pt}\discretionary{.}{%
}{.}\hspace{.4pt}2011\hspace{.1pt}\discretionary{.}{%
}{.}\hspace{.4pt}188}}


\bibitem{kretzschmar2023visual}
V.~Kretzschmar, G.~Scheuermann, M.~Stommel, and C.~Gillmann.
\newblock A visual analytics inspired approach to correlate and understand multiple mechanical tensor fields.
\newblock In {\em 2023 IEEE 16th Pacific Visualization Symposium (PacificVis)}, pp. 107--111, 2023. doi: {{%
10\hspace{.1pt}\discretionary{.}{%
}{.}\hspace{.4pt}1109\discretionary{/}{%
}{/}PacificVis56936\hspace{.1pt}\discretionary{.}{%
}{.}\hspace{.4pt}2023\hspace{.1pt}\discretionary{.}{%
}{.}\hspace{.4pt}00019}}


\bibitem{maack2024workflow}
R.~G. Maack, F.~Raith, J.~F. P{\'e}rez, G.~Scheuermann, and C.~Gillmann.
\newblock A workflow to systematically design uncertainty-aware visual analytics applications.
\newblock {\em The Visual Computer}, pp. 1--14, 2024. doi: {{%
10\hspace{.1pt}\discretionary{.}{%
}{.}\hspace{.4pt}1007\discretionary{/}{%
}{/}s00371\discretionary{%
}{-}{-}024\discretionary{%
}{-}{-}03435\discretionary{%
}{-}{-}x}}


\bibitem{meuschke2017glyph}
M.~Meuschke, S.~Voß, O.~Beuing, B.~Preim, and K.~Lawonn.
\newblock Glyph-based comparative stress tensor visualization in cerebral aneurysms.
\newblock In {\em Computer Graphics Forum}, vol.~36, pp. 99--108, 2017. doi: {{%
10\hspace{.1pt}\discretionary{.}{%
}{.}\hspace{.4pt}1111\discretionary{/}{%
}{/}cgf\hspace{.1pt}\discretionary{.}{%
}{.}\hspace{.4pt}13171}}


\bibitem{mirzargar2014curve}
M.~Mirzargar, R.~T. Whitaker, and R.~M. Kirby.
\newblock Curve boxplot: Generalization of boxplot for ensembles of curves.
\newblock {\em IEEE Transactions on Visualization and Computer Graphics}, 20(12):2654--2663, 2014. doi: {{%
10\hspace{.1pt}\discretionary{.}{%
}{.}\hspace{.4pt}1109\discretionary{/}{%
}{/}TVCG\hspace{.1pt}\discretionary{.}{%
}{.}\hspace{.4pt}2014\hspace{.1pt}\discretionary{.}{%
}{.}\hspace{.4pt}2346455}}


\bibitem{oster2018core}
T.~Oster, C.~R{\"o}ssl, and H.~Theisel.
\newblock Core lines in 3d second-order tensor fields.
\newblock In {\em Computer Graphics Forum (Proc. EuroVis)}, vol.~37, pp. 327--337, 2018. doi: {{%
10\hspace{.1pt}\discretionary{.}{%
}{.}\hspace{.4pt}1111\discretionary{/}{%
}{/}cgf\hspace{.1pt}\discretionary{.}{%
}{.}\hspace{.4pt}13423}}


\bibitem{oster2018parallel}
T.~Oster, C.~R{\"o}ssl, and H.~Theisel.
\newblock The parallel eigenvectors operator.
\newblock In {\em Proc. of Vision, Modeling, and Visualization (VMV 2018)}, pp. 39--46, 2018. doi: {{%
10\hspace{.1pt}\discretionary{.}{%
}{.}\hspace{.4pt}2312\discretionary{/}{%
}{/}vmv\hspace{.1pt}\discretionary{.}{%
}{.}\hspace{.4pt}20181251}}


\bibitem{padilla2020uncertainty}
L.~Padilla, M.~Kay, and J.~Hullman.
\newblock Uncertainty visualization.
\newblock {\em Wiley StatsRef: Statistics Reference Online}, pp. 1--18, 2021. doi: {{%
10\hspace{.1pt}\discretionary{.}{%
}{.}\hspace{.4pt}1002\discretionary{/}{%
}{/}9781118445112\hspace{.1pt}\discretionary{.}{%
}{.}\hspace{.4pt}stat08296}}


\bibitem{palacios2015feature}
J.~Palacios, H.~Yeh, W.~Wang, Y.~Zhang, R.~S. Laramee, R.~Sharma, T.~Schultz, and E.~Zhang.
\newblock Feature surfaces in symmetric tensor fields based on eigenvalue manifold.
\newblock {\em IEEE Transactions on Visualization and Computer Graphics}, 22(3):1248--1260, 2016. doi: {{%
10\hspace{.1pt}\discretionary{.}{%
}{.}\hspace{.4pt}1109\discretionary{/}{%
}{/}TVCG\hspace{.1pt}\discretionary{.}{%
}{.}\hspace{.4pt}2015\hspace{.1pt}\discretionary{.}{%
}{.}\hspace{.4pt}2484343}}


\bibitem{pothkow2012uncertainty}
K.~P{\"o}thkow and H.-C. Hege.
\newblock Uncertainty propagation in dt-mri anisotropy isosurface extraction.
\newblock In {\em New Developments in the Visualization and Processing of Tensor Fields}, pp. 209--225. Springer Berlin Heidelberg, Berlin, Heidelberg, 2012.

\bibitem{potter2012quantification}
K.~Potter, P.~Rosen, and C.~R. Johnson.
\newblock From quantification to visualization: A taxonomy of uncertainty visualization approaches.
\newblock In {\em Uncertainty Quantification in Scientific Computing}, pp. 226--249. Springer Berlin Heidelberg, Berlin, Heidelberg, 2012.

\bibitem{potter2009ensemble}
K.~Potter, A.~Wilson, P.-T. Bremer, D.~Williams, C.~Doutriaux, V.~Pascucci, and C.~R. Johnson.
\newblock Ensemble-vis: A framework for the statistical visualization of ensemble data.
\newblock In {\em 2009 IEEE International Conference on Data Mining Workshops}, pp. 233--240, 2009. doi: {{%
10\hspace{.1pt}\discretionary{.}{%
}{.}\hspace{.4pt}1109\discretionary{/}{%
}{/}ICDMW\hspace{.1pt}\discretionary{.}{%
}{.}\hspace{.4pt}2009\hspace{.1pt}\discretionary{.}{%
}{.}\hspace{.4pt}55}}


\bibitem{qu2020mode}
B.~Qu, L.~Roy, Y.~Zhang, and E.~Zhang.
\newblock Mode surfaces of symmetric tensor fields: Topological analysis and seamless extraction.
\newblock {\em IEEE Transactions on Visualization and Computer Graphics}, 27(2):583--592, 2021. doi: {{%
10\hspace{.1pt}\discretionary{.}{%
}{.}\hspace{.4pt}1109\discretionary{/}{%
}{/}TVCG\hspace{.1pt}\discretionary{.}{%
}{.}\hspace{.4pt}2020\hspace{.1pt}\discretionary{.}{%
}{.}\hspace{.4pt}3030431}}


\bibitem{roy2018robust}
L.~Roy, P.~Kumar, Y.~Zhang, and E.~Zhang.
\newblock Robust and fast extraction of 3d symmetric tensor field topology.
\newblock {\em IEEE Transactions on Visualization and Computer Graphics}, 25(1):1102--1111, 2019. doi: {{%
10\hspace{.1pt}\discretionary{.}{%
}{.}\hspace{.4pt}1109\discretionary{/}{%
}{/}TVCG\hspace{.1pt}\discretionary{.}{%
}{.}\hspace{.4pt}2018\hspace{.1pt}\discretionary{.}{%
}{.}\hspace{.4pt}2864768}}


\bibitem{schroeder2004visualization}
W.~Schroeder, K.~Martin, and B.~Lorensen.
\newblock {\em {The Visualization Toolkit (4th ed.)}}.
\newblock Kitware, 2006.

\bibitem{schultz2011topological}
T.~Schultz.
\newblock Topological features in 2d symmetric higher-order tensor fields.
\newblock In {\em Computer Graphics Forum}, vol.~30, pp. 841--850, 2011. doi: {{%
10\hspace{.1pt}\discretionary{.}{%
}{.}\hspace{.4pt}1111\discretionary{/}{%
}{/}j\hspace{.1pt}\discretionary{.}{%
}{.}\hspace{.4pt}1467\discretionary{%
}{-}{-}8659\hspace{.1pt}\discretionary{.}{%
}{.}\hspace{.4pt}2011\hspace{.1pt}\discretionary{.}{%
}{.}\hspace{.4pt}01933\hspace{.1pt}\discretionary{.}{%
}{.}\hspace{.4pt}x}}


\bibitem{schultz2010superquadric}
T.~Schultz and G.~L. Kindlmann.
\newblock Superquadric glyphs for symmetric second-order tensors.
\newblock {\em IEEE Transactions on Visualization and Computer Graphics}, 16(6):1595--1604, 2010. doi: {{%
10\hspace{.1pt}\discretionary{.}{%
}{.}\hspace{.4pt}1109\discretionary{/}{%
}{/}TVCG\hspace{.1pt}\discretionary{.}{%
}{.}\hspace{.4pt}2010\hspace{.1pt}\discretionary{.}{%
}{.}\hspace{.4pt}199}}


\bibitem{schultz2013hifive}
T.~Schultz, L.~Schlaffke, B.~Schölkopf, and T.~Schmidt-Wilcke.
\newblock Hifive: A hilbert space embedding of fiber variability estimates for uncertainty modeling and visualization.
\newblock In {\em Computer Graphics Forum}, vol.~32, pp. 121--130, 2013. doi: {{%
10\hspace{.1pt}\discretionary{.}{%
}{.}\hspace{.4pt}1111\discretionary{/}{%
}{/}cgf\hspace{.1pt}\discretionary{.}{%
}{.}\hspace{.4pt}12099}}


\bibitem{schultz2014fuzzy}
T.~Schultz, A.~Vilanova, R.~Brecheisen, and G.~Kindlmann.
\newblock Fuzzy fibers: Uncertainty in dmri tractography.
\newblock {\em Scientific Visualization: Uncertainty, Multifield, Biomedical, and Scalable Visualization}, pp. 79--92, 2014. doi: {{%
10\hspace{.1pt}\discretionary{.}{%
}{.}\hspace{.4pt}1007\discretionary{/}{%
}{/}978\discretionary{%
}{-}{-}1\discretionary{%
}{-}{-}4471\discretionary{%
}{-}{-}6497\discretionary{%
}{-}{-}5\_8}}


\bibitem{siddiqui2021uncertainty}
F.~Siddiqui, T.~H{\"o}llt, and A.~Vilanova.
\newblock Uncertainty in the dti visualization pipeline.
\newblock In {\em Anisotropy Across Fields and Scales}, pp. 125--148. Springer International Publishing, Cham, 2021.

\bibitem{siddiqui2021progressive}
F.~Siddiqui, T.~Höllt, and A.~Vilanova.
\newblock A progressive approach for uncertainty visualization in diffusion tensor imaging.
\newblock In {\em Computer Graphics Forum}, pp. 411--422, 2021. doi: {{%
10\hspace{.1pt}\discretionary{.}{%
}{.}\hspace{.4pt}1111\discretionary{/}{%
}{/}cgf\hspace{.1pt}\discretionary{.}{%
}{.}\hspace{.4pt}14317}}


\bibitem{tricoche2008invariant}
X.~Tricoche, G.~Kindlmann, and C.-F. Westin.
\newblock Invariant crease lines for topological and structural analysis of tensor fields.
\newblock {\em IEEE Transactions on Visualization and Computer Graphics}, 14(6):1627--1634, 2008. doi: {{%
10\hspace{.1pt}\discretionary{.}{%
}{.}\hspace{.4pt}1109\discretionary{/}{%
}{/}TVCG\hspace{.1pt}\discretionary{.}{%
}{.}\hspace{.4pt}2008\hspace{.1pt}\discretionary{.}{%
}{.}\hspace{.4pt}148}}


\bibitem{wang2018visualization}
J.~Wang, S.~Hazarika, C.~Li, and H.-W. Shen.
\newblock Visualization and visual analysis of ensemble data: A survey.
\newblock {\em IEEE Transactions on Visualization and Computer Graphics}, 25(9):2853--2872, 2019. doi: {{%
10\hspace{.1pt}\discretionary{.}{%
}{.}\hspace{.4pt}1109\discretionary{/}{%
}{/}TVCG\hspace{.1pt}\discretionary{.}{%
}{.}\hspace{.4pt}2018\hspace{.1pt}\discretionary{.}{%
}{.}\hspace{.4pt}2853721}}


\bibitem{weiskopf2022uncertainty}
D.~Weiskopf.
\newblock Uncertainty visualization: Concepts, methods, and applications in biological data visualization.
\newblock {\em Frontiers in Bioinformatics}, 2, 2022. doi: {{%
10\hspace{.1pt}\discretionary{.}{%
}{.}\hspace{.4pt}3389\discretionary{/}{%
}{/}fbinf\hspace{.1pt}\discretionary{.}{%
}{.}\hspace{.4pt}2022\hspace{.1pt}\discretionary{.}{%
}{.}\hspace{.4pt}793819}}


\bibitem{whitaker2013contour}
R.~T. Whitaker, M.~Mirzargar, and R.~M. Kirby.
\newblock Contour boxplots: A method for characterizing uncertainty in feature sets from simulation ensembles.
\newblock {\em IEEE Transactions on Visualization and Computer Graphics}, 19(12):2713--2722, 2013. doi: {{%
10\hspace{.1pt}\discretionary{.}{%
}{.}\hspace{.4pt}1109\discretionary{/}{%
}{/}TVCG\hspace{.1pt}\discretionary{.}{%
}{.}\hspace{.4pt}2013\hspace{.1pt}\discretionary{.}{%
}{.}\hspace{.4pt}143}}


\bibitem{zhang2017overview+}
C.~Zhang, M.~W. Caan, T.~H{\"o}llt, E.~Eisemann, and A.~Vilanova.
\newblock Overview+ detail visualization for ensembles of diffusion tensors.
\newblock In {\em Computer Graphics Forum}, vol.~36, pp. 121--132. Wiley Online Library, 2017. doi: {{%
10\hspace{.1pt}\discretionary{.}{%
}{.}\hspace{.4pt}1111\discretionary{/}{%
}{/}cgf\hspace{.1pt}\discretionary{.}{%
}{.}\hspace{.4pt}13173}}


\bibitem{zhang2017comparative}
C.~Zhang, T.~Höllt, M.~W.~A. Caan, E.~Eisemann, and A.~Vilanova.
\newblock {Comparative Visualization for Diffusion Tensor Imaging Group Study at Multiple Levels of Detail}.
\newblock In {\em Eurographics Workshop on Visual Computing for Biology and Medicine}. The Eurographics Association, 2017. doi: {{%
10\hspace{.1pt}\discretionary{.}{%
}{.}\hspace{.4pt}2312\discretionary{/}{%
}{/}vcbm\hspace{.1pt}\discretionary{.}{%
}{.}\hspace{.4pt}20171237}}


\bibitem{zhang2015glyph}
C.~Zhang, T.~Schultz, K.~Lawonn, E.~Eisemann, and A.~Vilanova.
\newblock Glyph-based comparative visualization for diffusion tensor fields.
\newblock {\em IEEE Transactions on Visualization and Computer Graphics}, 22(1):797–806, jan 2016. doi: {{%
10\hspace{.1pt}\discretionary{.}{%
}{.}\hspace{.4pt}1109\discretionary{/}{%
}{/}TVCG\hspace{.1pt}\discretionary{.}{%
}{.}\hspace{.4pt}2015\hspace{.1pt}\discretionary{.}{%
}{.}\hspace{.4pt}2467435}}


\bibitem{zhang2022enconvis}
M.~Zhang, Q.~Li, L.~Chen, X.~Yuan, and J.~Yong.
\newblock Enconvis: A unified framework for ensemble contour visualization.
\newblock {\em IEEE Transactions on Visualization and Computer Graphics}, 29(4):2067--2079, 2023. doi: {{%
10\hspace{.1pt}\discretionary{.}{%
}{.}\hspace{.4pt}1109\discretionary{/}{%
}{/}TVCG\hspace{.1pt}\discretionary{.}{%
}{.}\hspace{.4pt}2021\hspace{.1pt}\discretionary{.}{%
}{.}\hspace{.4pt}3140153}}


\bibitem{zhang2017applying}
Y.~Zhang, X.~Gao, and E.~Zhang.
\newblock Applying 2d tensor field topology to solid mechanics simulations.
\newblock In {\em Modeling, Analysis, and Visualization of Anisotropy}, pp. 29--41. Springer International Publishing, Cham, 2017.

\bibitem{zhang2021degenerate}
Y.~Zhang, H.~Nie, and E.~Zhang.
\newblock Degenerate curve bifurcations in 3d linear symmetric tensor fields.
\newblock In {\em Anisotropy Across Fields and Scales}, pp. 23--38. Springer International Publishing, Cham, 2021.

\bibitem{zheng2005topological}
X.~Zheng, B.~Parlett, and A.~Pang.
\newblock Topological lines in 3d tensor fields and discriminant hessian factorization.
\newblock {\em IEEE Transactions on Visualization and Computer Graphics}, 11(4):395--407, 2005. doi: {{%
10\hspace{.1pt}\discretionary{.}{%
}{.}\hspace{.4pt}1109\discretionary{/}{%
}{/}TVCG\hspace{.1pt}\discretionary{.}{%
}{.}\hspace{.4pt}2005\hspace{.1pt}\discretionary{.}{%
}{.}\hspace{.4pt}67}}


\bibitem{zheng2006degenerate}
X.~Zheng, X.~Tricoche, and A.~Pang.
\newblock Degenerate 3d tensors.
\newblock {\em Visualization and Processing of Tensor Fields}, pp. 241--256, 2006. doi: {{%
10\hspace{.1pt}\discretionary{.}{%
}{.}\hspace{.4pt}1007\discretionary{/}{%
}{/}3\discretionary{%
}{-}{-}540\discretionary{%
}{-}{-}31272\discretionary{%
}{-}{-}2\_14}}


\bibitem{zobel2018extremal}
V.~Zobel and G.~Scheuermann.
\newblock Extremal curves and surfaces in symmetric tensor fields.
\newblock {\em Vis. Comput.}, 34(10):1427–1442, oct 2018. doi: {{%
10\hspace{.1pt}\discretionary{.}{%
}{.}\hspace{.4pt}1007\discretionary{/}{%
}{/}s00371\discretionary{%
}{-}{-}017\discretionary{%
}{-}{-}1450\discretionary{%
}{-}{-}1}}


\end{thebibliography}
\end{document}